\documentclass[iop]{emulateapj}
\usepackage{graphicx}
\usepackage{amsmath}
\usepackage{natbib}
\usepackage[pdftex,colorlinks,citecolor=blue,linkcolor=blue,pdfauthor=Julian~Merten~&~CLASH,pdftitle=CLASH~c-M~Relation,pdfsubject=Cosmology]{hyperref}

\providecommand{\eprint}[1]{\href{http://arxiv.org/abs/#1}{#1}}

\providecommand{\adsurl}[1]{\href{#1}{ADS}}

\begin{document}

\shorttitle{CLASH: The c-M relation}
\shortauthors{J.~Merten \& The CLASH collaboration}
\slugcomment{Submitted to the Astrophysical Journal \textcopyright~2014. All rights reserved}

\title{CLASH: The concentration-Mass relation of galaxy clusters}

\author{J. Merten\altaffilmark{1,2}}
\author{M. Meneghetti\altaffilmark{1,3,4}}
\author{M. Postman\altaffilmark{5}}
\author{K. Umetsu\altaffilmark{6}}
\author{A. Zitrin\altaffilmark{2,7}}
\author{E. Medezinski\altaffilmark{8}}
\author{M. Nonino\altaffilmark{9}}
\author{A. Koekemoer\altaffilmark{5}}
\author{P. Melchior\altaffilmark{10}}
\author{D. Gruen\altaffilmark{11,12}}
\author{L.A. Moustakas\altaffilmark{1}}
\author{M. Bartelmann\altaffilmark{13}}
\author{O. Host\altaffilmark{14}}
\author{M. Donahue\altaffilmark{15}}
\author{D. Coe\altaffilmark{5}}
\author{A. Molino\altaffilmark{16}}
\author{S. Jouvel\altaffilmark{17,18}}
\author{A. Monna\altaffilmark{11,12}}
\author{S. Seitz\altaffilmark{11,12}}
\author{N. Czakon\altaffilmark{6}}
\author{D. Lemze\altaffilmark{8}}
\author{J. Sayers\altaffilmark{2}}
\author{I. Balestra\altaffilmark{9,19}}
\author{P. Rosati\altaffilmark{20}}
\author{N. Ben\'itez\altaffilmark{15}}
\author{A. Biviano\altaffilmark{9}}
\author{R. Bouwens\altaffilmark{21}}
\author{L. Bradley\altaffilmark{5}}
\author{T. Broadhurst\altaffilmark{22,23}}
\author{M. Carrasco\altaffilmark{24,13}}
\author{H. Ford\altaffilmark{8}}
\author{C. Grillo\altaffilmark{14}}
\author{L. Infante\altaffilmark{25}}
\author{D. Kelson\altaffilmark{26}}
\author{O. Lahav\altaffilmark{17}}
\author{R. Massey\altaffilmark{27}}
\author{J. Moustakas\altaffilmark{28}}
\author{E. Rasia\altaffilmark{29}}
\author{J. Rhodes\altaffilmark{1,2}}
\author{J. Vega\altaffilmark{30,31}}
\author{W. Zheng\altaffilmark{5}}
 
\email{jmerten@caltech.edu} 
\altaffiltext{1}{Jet Propulsion Laboratory, California Institute of Technology, 4800 Oak Grove Drive, Pasadena, CA 91109, USA}
\altaffiltext{2}{California Institute of Technology, MC 249-17, Pasadena, CA 91125, USA}
\altaffiltext{3}{INAF, Osservatorio Astronomico di Bologna, via Ranzani 1, 40127 Bologna, Italy}
\altaffiltext{4}{INFN, Sezione di Bologna, Viale Berti Pichat 6/2, 40127 Bologna, Italy}
\altaffiltext{5}{Space Telescope Science Institute, 3700 San Martin Drive, Baltimore, MD 21208, USA}
\altaffiltext{6}{Institute of Astronomy and Astrophysics, Academia Sinica, P.O. Box 23-141, Taipei 10617, Taiwan}
\altaffiltext{7}{Hubble Fellow}
\altaffiltext{8}{Department of Physics and Astronomy, The Johns Hopkins University, 3400 North Charles Street, Baltimore, MD 21218, USA}
\altaffiltext{9}{INAF/Osservatorio Astronomico di Trieste, via G.B. Tiepolo 11, I-34143 Trieste, Italy}
\altaffiltext{10}{Center for Cosmology and Astro-Particle Physics and Department of Physics, The Ohio State University, Columbus, OH 43210, USA}
\altaffiltext{11}{Universit\"{a}ts-Sternwarte M\"{u}nchen, Scheinerstr. 1, D-81679 M\"{u}nchen, Germany}
\altaffiltext{12}{Max-Planck-Institute f\"ur extraterrestrische Physik, Giessenbachstr. 1, 85748 Garching, Germany}
\altaffiltext{13}{Universit\"{a}t Heidelberg, Zentrum f\"{u}r Astronomie, Institut f\"{u}r Theoretische Astrophysik, Philosophenweg 12, 69120 Heidelberg, Germany}
\altaffiltext{14}{Dark Cosmology Centre, Niels Bohr Institute, University of Copenhagen, Juliane Maries Vej 30, DK-2100 Copenhagen, Denmark}
\altaffiltext{15}{Department of Physics and Astronomy, Michigan State University, East Lansing, MI 48824, USA}
\altaffiltext{16}{Instituto de Astrof\'{\i}sica de Andaluc\'{\i}a (CSIC), E-18080 Granada, Spain}
\altaffiltext{17}{Institut de Ci\'{e}ncies de l'Espai (IEEC-CSIC), E-08193 Bellaterra (Barcelona), Spain}
\altaffiltext{18}{Department of Physics and Astronomy, University College London, London WC1E 6BT, UK}
\altaffiltext{19}{INAF - Osservatorio Astronomico di Capodimonte, Via Moiariello 16, I-80131 Napoli, Italy}
\altaffiltext{20}{Dipartimento di Fisica e Scienze della Terra, Universit\`{a} degli Studi di Ferrara, Via Saragat 1, I-44122 Ferrara, Italy}
\altaffiltext{21}{Leiden Observatory, Leiden University, P. O. Box 9513, NL-2333 Leiden, The Netherlands}
\altaffiltext{22}{Department of Theoretical Physics and History of Science, University of the Basque Country UPV/EHU, P.O. Box 644, E-48080 Bilbao, Spain}
\altaffiltext{23}{Ikerbasque, Basque Foundation for Science, Alameda Urquijo, 36-5 Plaza Bizkaia, E-48011 Bilbao, Spain}
\altaffiltext{24}{Instituto de Astrofísica, Facultad de F\'{\i}sica, Pontificia Universidad Cat\'{o}lica de Chile, Casilla 306, Santiago 22, Chile}
\altaffiltext{25}{Centro de Astro-Ingenier\'{\i}a, Departamento de Astronom\'{\i}a y Astrof\'{\i}sica, Pontificia Universidad Catolica de Chile, V. Mackenna 4860, Santiago, Chile}
\altaffiltext{26}{Observatories of the Carnegie Institution of Washington, Pasadena, CA 91101, USA}
\altaffiltext{27}{Institute for Computational Cosmology, Durham University, South Road, Durham DH1 3LE, UK}
\altaffiltext{28}{Department of Physics and Astronomy, Siena College, 515 Loudon Road, Loudonville, NY 12211, USA}
\altaffiltext{29}{Physics Dept., University of Michigan, 450 Church Ave, Ann Arbor, MI 48109, USA}
\altaffiltext{30}{Departamento de F\'{\i}sica Te\'{o}rica, Universidad Aut\'{o}noma de Madrid, Cantoblanco, E-28049 Madrid, Spain}
\altaffiltext{31}{LERMA, CNRS UMR 8112, Observatoire de Paris, 61 Avenue de l'Observatoire, 75014 Paris, France}

\begin{abstract}
We present a new determination of the concentration-mass relation for galaxy clusters based on 
our comprehensive lensing analysis of 19 X-ray selected galaxy clusters 
from the Cluster Lensing and Supernova Survey with Hubble (CLASH). Our sample spans a redshift range between 0.19 and
0.89. We combine weak-lensing constraints from the
Hubble Space Telescope (HST) and from ground-based wide-field data with strong lensing constraints from HST. 
The result are reconstructions of the surface-mass density for all CLASH clusters on multi-scale grids.
Our derivation of NFW parameters yields virial masses between $0.53\times10^{15} M_{\odot}/h$ and $1.76\times10^{15} M_{\odot}/h$
and the halo concentrations are distributed around $c_{200\textrm{c}}\sim3.7$ with a $1\sigma$ significant negative trend with cluster mass.
We find an excellent 4\% agreement between 
our measured concentrations and the expectation from numerical simulations
after accounting for the CLASH selection function based on X-ray morphology. The simulations are analyzed
in 2D to account for possible biases in the lensing reconstructions due to projection effects. 
The theoretical concentration-mass (c-M) relation from our X-ray selected set of simulated clusters and the c-M relation 
derived directly from the CLASH data agree at the 90\% confidence level.
\end{abstract}


\section{Introduction}\label{intro}
The standard model of cosmology ($\Lambda$CDM) is extremely successful in explaining the observed large-scale structure of the Universe \citep[see e.g.][]{PlanckCollaboration2013, Anderson2012}. 
However, when moving to progressively smaller length scales, inconsistencies  between theoretical predictions and real observations have emerged. 
Examples include the cored mass-density profiles of dwarf-spheroidal galaxies \citep{Walker2011}, the abundance of Milky Way satellites \citep{Boylan-Kolchin2012} and the flat dark matter density profiles in the cores of galaxy clusters \citep{Sand2002,Newman2013}. 

Galaxy clusters are unique tracers of cosmological structure formation \citep[e.g.][]{Voit2005,Borgani2011}. As the largest collapsed objects in the observable Universe, clusters form the bridge between the large-scale structure of the Universe and the astrophysical regime of individual halos.  
From an observational point of view, all main mass components of a cluster, hot ionized gas, dark matter and luminous stars, are directly or indirectly observable with the help of X-ray observatories \citep[e.g.][]{Rosati2002,Ettori2013}, gravitational lensing \citep[e.g.][]{Bartelmann2001,Bartelmann2010a} or optical observations.  	 

As shown by numerical simulations \citep{Navarro1996}, dark matter tends to arrange itself following a specific, spherically symmetric density profile
\begin{equation}\label{NFW}
\rho_{\mathrm{NFW}}(r) = \frac{\rho_{\mathrm{s}}}{r/r_{\mathrm{s}}\left(1+r/r_{\mathrm{s}}\right)^{2}},
\end{equation}
where the only two parameters $\rho_{\mathrm{s}}$ and $r_{\mathrm{s}}$ are a scale density and a scale radius. This functional form is now commonly called the Navarro, Frenk and White (NFW) density profile. It was found to fit well the dark matter distribution of halos in numerical simulations, independent of halo mass, cosmological parameters or formation time \citep{Navarro1997,Bullock2001}. 

A specific parametrization of the NFW profile uses the total mass enclosed within a certain radius $r_{\Delta}$
\begin{equation}\label{NFW_mass}
M_{\Delta} = 4\pi\rho_{\mathrm{s}}r_{\mathrm{s}}^{3}\left(\ln\left(1+c_{\Delta}\right)-\frac{c_{\Delta}}{1+c_{\Delta}}\right),
\end{equation}
and the concentration parameter
\begin{equation}\label{concentration}
 c_{\Delta} = \frac{r_{\Delta}}{r_{s}}.
\end{equation}
When applying the relations above  to a specific analysis, the radius $r_{\Delta}$ is chosen such that it describes the halo on the scale
of interest. An example is the radius at which the average density of the halo is 200 times the critical density of the Universe
at this redshift ($\Delta=200c$).
Cosmological simulations show that dark matter structures occupy a specific region in the concentration-mass plane. This defines the concentration-mass (c-M) relation which is a mild function of formation redshift and halo mass \citep{Bullock2001,Eke2001a,Zhao2003a,Gao2008,Duffy2008,Klypin2011,Prada2012,Bhattacharya2013}. 

Observational efforts have been undertaken to measure the c-M relation either using gravitational lensing \citep{Comerford2007,Oguri2012,Okabe2013}, X-ray observations \citep{Buote2007,Schmidt2007,Ettori2010} or dynamical 
analysis of cluster members \citep{Lemze2009,Wojtak2010,Biviano2013}. Some of the observed relations are in tension with the predictions of numerical simulations \citep{Fedeli2012,Duffy2008}.
The most prominent example of such tension is the cluster Abell 1689 \citep[][and references therein]{Broadhurst2005,Peng2009}, 
with a concentration parameter up to a factor of three higher than predicted.
In a follow-up study, \citet{Broadhurst2008} compared a larger sample of five clusters to the prediction from $\Lambda$CDM and found the derived c-M relation in tension with the theoretical expectations \citep[see also][]{Broadhurst2008a,Zitrin2010a,Meneghetti2011}.
Possible explanations for these discrepancies include a selection-bias of the cluster sample since these clusters were known strong lenses, paired with the assumption of spherical symmetry for these systems \citep{Hennawi2007,Meneghetti2010}. Moreover, the influence of baryons on the cluster core \citep{Fedeli2012,Killedar2012} and even the effects of early dark energy \citep{Fedeli2007,Sadeh2008,Francis2009,Grossi2009} have been introduced as possible explanations.
Ultimately, a new set of high-quality observations of an unbiased ensemble of clusters was needed to answer the question if observed galaxy clusters are indeed in tension with our cosmological standard model. 

The Cluster Lensing And Supernova Survey with Hubble (CLASH) \citep{Postman2012} is a multi-cycle treasury program, using 524 Hubble Space Telescope (HST) orbits to target 25 galaxy clusters, largely drawn from the Abell and MACS cluster catalogs \citep{Abell1958,Abell1989,Ebeling2001,Ebeling2007,Ebeling2010}. 
Twenty clusters were specifically selected by their largely unperturbed X-ray morphology with the goal of representing a sample of clusters with regular, unbiased density profiles that allow for an optimal comparison to models of cosmological structure formation.
As reported in \citet{Postman2012} all clusters of the sample are fairly X-ray luminous with X-ray temperatures $T_{\mathrm{x}}\geq 5$ keV and show a smooth morphology in their X-ray surface brightness. For all systems the separation 
between the brightest cluster galaxy (BCG) and the X-ray luminosity centroid is $<20$~kpc.
An overview of the basic properties of the sample can be found in Table~\ref{CLASH_CLUSTERS}.  
In the following we will use these X-ray selected clusters to derive the observed c-M relation for CLASH clusters based on weak and strong lensing and perform a thorough comparison to the theoretical expectation from numerical simulations.
This study has two companion papers. The weak-lensing and magnification analysis of CLASH clusters by \citet{Umetsu2014} and the detailed characterization of numerical simulations of CLASH clusters by \citet{Meneghetti2014}.

This paper is structured as follows: Sec.~\ref{Theory} provides a basic introduction to gravitational lensing and introduces the method used to recover the dark matter distribution from the observational data. The respective input data is described in Sec.~\ref{Data} and the resulting mass maps and density profiles of the CLASH clusters are presented in Sec.~\ref{Results}. We interpret our results by a detailed comparison to theoretical c-M relations from the literature in Sec.~\ref{GMc} and use our own tailored set of simulations to derive a CLASH-like c-M relation in Sec.~\ref{SMc}. We conclude in Sec.~\ref{Concl}. Throughout this work we assume a flat cosmological model similar to a WMAP7 cosmology \citep{Komatsu2011} with $\Omega_{\textrm{m}} = 0.27$,  $\Omega_{\Lambda} = 0.73$ and a Hubble constant of $h = 0.7$. For the redshift range of our cluster sample this translates to physical distance scales of 3.156 -- 7.897 kpc/\arcsec.  

\begin{deluxetable*}{lcccccc}

 \tablecaption{The CLASH X-ray selected cluster sample\label{CLASH_CLUSTERS}}
 \tablewidth{0pt}
 \tablehead{
   \colhead{Name} & \colhead{z} & \colhead{R.A.} & \colhead{DEC}& \colhead{k$T_{\mathrm{X}}$\tablenotemark{a}}&\colhead{$L_{\mathrm{bol}}$\tablenotemark{a}}& \colhead{$\arcsec\rightarrow$kpc\tablenotemark{b}} \\
   &&\colhead{[deg/J2000]} &\colhead{[deg/J2000]}&\colhead{[keV]}&\colhead{[$10^{44}$erg/s]}&
   }
\startdata
Abell~383&	0.188&	42.014090&     -3.5292641&	6.5&	6.7&	3.156\\
Abell~209&	0.206&	22.968952&     -13.611272&	7.3&	12.7&	3.392\\
Abell~1423&	0.213&	179.32234&	33.610973&	7.1&	7.8&	3.482\\
Abell~2261&	0.225&	260.61336&	32.132465&	7.6&	18.0&	3.632\\
RXJ2129+0005&	0.234&	322.41649&	0.0892232&	5.8&	11.4&	3.742\\
Abell~611&	0.288&	120.23674&	36.056565&	7.9&	11.7&	4.357\\
MS2137-2353&	0.313&	325.06313&     -23.661136&	5.9&	9.9&	4.617\\
RXCJ2248-4431&	0.348&	342.18322&     -44.530908&	12.4&	69.5&	4.959\\
MACSJ1115+0129&	0.352&	168.96627&	1.4986116&	8.0&	21.1&	4.996\\
MACSJ1931-26&	0.352&	292.95608&     -26.575857&	6.7&	20.9&	4.996\\
RXJ1532.8+3021&	0.363&	233.22410&	30.349844&	5.5&	20.5&	4.931\\
MACSJ1720+3536&	0.391&	260.06980&	35.607266&	6.6&	13.3&	5.343\\
MACSJ0429-02&	0.399&	67.400028&     -2.8852066&	6.0&	11.2&	5.411\\
MACSJ1206-08&	0.439&	181.55065&     -8.8009395&	10.8&	43.0&	5.732\\
MACSJ0329-02&	0.450&	52.423199&     -2.1962279&	8.0&	17.0&	5.815\\
RXJ1347-1145&	0.451&	206.87756&     -11.752610&	15.5&	90.8&	5.822\\
MACSJ1311-03&	0.494&	197.75751&     -3.1777029&	5.9&	9.4&	6.128\\
MACSJ1423+24&	0.545&	215.94949&	24.078459&	6.5&	14.5&	6.455\\
MACSJ0744+39&	0.686&	116.22000&	39.457408&	8.9&	29.1&	7.186\\
CLJ1226+3332&	0.890&	186.74270&	33.546834&	13.8&	34.4&	7.897\\
\enddata
\tablenotetext{a}{From \citet{Postman2012} and references therein.} 
\tablenotetext{b}{Conversion factor to convert arcseconds to kpc at the cluster's redshift and given the cosmological background model.}
  \end{deluxetable*}


\section{Cluster Mass Profiles from Gravitational Lensing}\label{Theory}
We use gravitational lensing to recover the distribution of matter in galaxy clusters from imaging data. Lensing is particularly well-suited for this purpose since it is sensitive to the lens' total matter content, independent of its composition and under a minimum number of assumptions. After we discussed the basics of this powerful technique we will present a non-parametric inversion algorithm which maps the dark matter mass distribution over a wide range of angular scales. 
The CLASH data were designed to provide a unique combination of angular resolution, depth and multi-wavelength coverage that allows many new multiply lensed galaxies to be identified and their redshifts to be accurately estimated. These data are ideal for use with the \texttt{SaWLens} algorithm,
which makes no \textit{a priori} assumptions about the distribution of matter in a galaxy cluster. 
\subsection{Gravitational Lensing}\label{Theor_Lensing}
Gravitational lensing is a direct consequence of Einstein's theory of general relativity \citep[see e.g.][for a complete derivation]{Bartelmann2010a}. For cluster-sized lenses the lens mapping can be described by the lens equation
\begin{equation}\label{lensequation}
\vec{\beta}= \vec{\theta}-\vec{\alpha}\left(\vec{\theta}\right).
\end{equation}
This lens equation describes how the original 2D angular position in the source plane $\vec{\beta} = (\beta_{1},\beta_{2})$ is shifted by a deflection angle $\vec{\alpha} = (\alpha_{1},\alpha_{2})$ to the angular coordinates $\vec{\theta} = (\theta_{1},\theta_{2})$ in the lens plane. From now on we denote the angular diameter distance between observer and lens as $D_{\mathrm{l}}$, between observer and source as $D_{\mathrm{s}}$, and between lens and source as $D_{\mathrm{ls}}$. The deflection angle depends on the surface-mass density distribution of the lens $\Sigma(D_{\mathrm{d}}\vec{\theta})$ and can be related to a lensing potential
\begin{equation}\label{lensingpotential}
  \psi(\vec{\theta}):= \frac{1}{\pi}\int d^{2}\theta^{'}\frac{\Sigma(D_{\mathrm{l}}\vec{\theta})}{\Sigma_{\mathrm{cr}}}\textrm{ln}|\vec{\theta}-\vec{\theta}^{'}|,
\end{equation}
which is a line-of-sight projected and rescaled version of the Newtonian potential. The cosmological background model enters this equation through the critical surface mass density for lensing given by
\begin{equation}\label{critical_dens}
\Sigma_{\mathrm{cr}}=\frac{c^{2}}{4\pi G}\frac{D_{\mathsf s}}{D_{\mathrm l}D_{\mathsf{ls}}},
\end{equation}
where $c$ is the speed of light and $G$ is Newton's constant. By introducing the complex lensing operators \citep[e.g.][]{Bacon2006,Schneider2008} $\partial := (\frac{\partial}{\partial\theta_{1}} +\mathrm{i}\frac{\partial}{\partial\theta_{2}})$ and $\partial^{*} := (\frac{\partial}{\partial\theta_{1}} -\mathrm{i}\frac{\partial}{\partial\theta_{2}})$ one can derive important lensing quantities as derivatives of the lensing potential
\begin{align}\label{lensingquantities}
 \alpha &:= \partial\psi \qquad \qquad  &s=1 \nonumber \\
 2\gamma &:=  \partial\partial \psi  &s=2 \\
 2\kappa &:= \partial \partial^{*} \psi  &s=0 \nonumber
\end{align}
where $\alpha$ is the complex form of the already known deflection angle, $\gamma$ is called the complex shear and the scalar quantity $\kappa$ is called convergence. The behavior of each quantity under rotations of the coordinate frame is given by the spin-parameter $s$. 

When relating these basic lens quantities to observables one distinguishes two specific regimes. In the case of weak lensing the distortions induced by the lens mapping are small and can be related to Eqs.~\ref{lensingquantities} by the reduced shear
\begin{equation}\label{reducedshear}
 g:=\frac{\gamma}{1-\kappa},
\end{equation}
which is directly proportional to measured complex ellipticities $\epsilon(\vec{\theta})$ of background sources in the lens' field. Due to the intrinsic ellipticity of galaxies, localized averages over an ensemble of sources are used to separate the lensing signal from the random orientation caused by the intrinsic ellipticity  
\begin{equation}
 g=\left<\epsilon\right>.
 \label{epsilon}
\end{equation}
For a more thorough description of the relation between the measured shapes of galaxy images and the lens properties in the weak lensing regime we refer to the review by \citet[][and references therein]{Bartelmann2001}. We also do not discuss here the many systematic effects to be taken into account during such a shape measurement but refer to e.g. \citet{Kitching2012,Massey2013}.

In the strong lensing regime, close to the core of the lens' mass distribution, the assumption of small image distortions does not hold any more. The lens equation becomes non-linear and therefore multiple images of the same source can form. 
This happens near the critical line at a given redshift which is defined by the roots of the lensing Jacobian
\begin{equation}\label{Jacobiandeterminant}
 \det\mathcal{A} = (1-\kappa)^{2}-\gamma^{2}.
\end{equation} 
While the weak lensing regime expands over the full cluster field, it does not describe the mass distribution in the center of the cluster. Strong lensing is limited to the inner-most $10\arcsec$ -- $50\arcsec$ of the cluster field, which renders the combination of the two regimes the ideal approach for mass reconstruction.

\subsection{Non-parametric Lensing Inversion with \texttt{SaWLens}}
\label{Theory_Sawlens}
\begin{figure}
 \includegraphics[width=.48\textwidth]{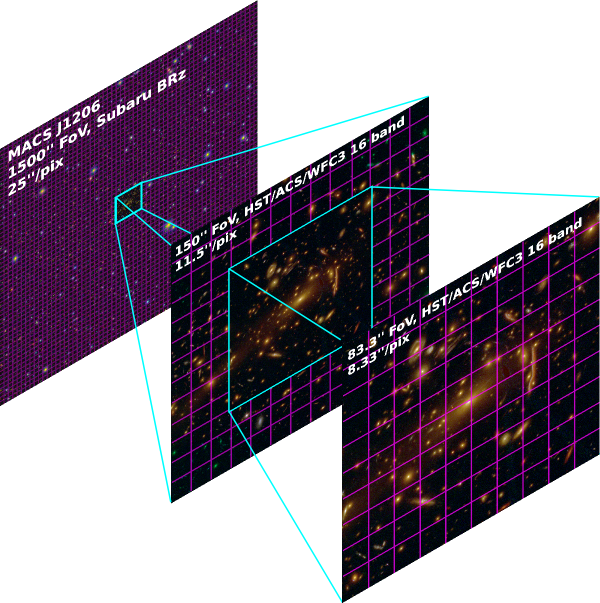}
 \caption{A visualization of our multi-scale approach. While weak lensing data from Subaru allows for a mass reconstruction of a galaxy cluster
 on a wide field, the achievable resolution is rather low. HST weak lensing delivers higher resolution but on a relatively
 small field. Finally, the strong lensing regime provides a very high resolution, but only in the inner-most cluster core. This figure shows one of our sample clusters, MACS~J1206 and the reconstruction grids for all three lensing regimes.
 \label{AdaptGrid}}
\end{figure}

The \texttt{SaWLens} (Strong -and Weak Lensing) method was developed with two goals in mind. First, it should consistently combine weak and strong lensing. The second goal was to make no \textit{a priori} assumptions about the underlying mass distribution, but to build solely upon the input data. The initial idea for such a reconstruction algorithm was formulated by \citet{Bartelmann1996} and was further developed by  \citet{Seitz1998} and \citet{Cacciato2006}. Similar ideas were implemented by \citet{Bradavc2005} with first applications
to observations in \citet{Bradavc2005a} and \citet{Bradav2006}.  
In its current implementation \citep{Merten2009}, \texttt{SaWLens} performs a reconstruction of the lensing potential (Eq.~\ref{lensingpotential}) on an adaptively refined grid. In this particular study, the method uses three different grid sizes to account for weak lensing 
on a wide field, such as is provided by ground-based telescopes, weak lensing constraints from the HST on a much smaller field-of-view but with considerably higher spatial resolution, and a fine grained grid to trace strong lensing features near the inner-most core of the cluster. This three-level adaptive grid is illustrated in Fig.~\ref{AdaptGrid}.

\texttt{SaWLens} uses a statistical approach to reconstruct the lensing potential $\psi$ in every pixel of the grid. A $\chi^{2}$-function, which depends on the lensing potential and includes a weak and a strong-lensing term  is defined by 
\begin{equation}
 \chi^{2}(\psi) = \chi^{2}_{\textrm{w}}(\psi)+\chi^{2}_{\textrm{s}}(\psi).
\label{overall_chi}
 \end{equation}
 and the algorithm minimizes it such that the input data is best described by a pixelized lensing potential $\psi_{l}$
 \begin{equation}
   \frac{\partial\chi^{2}(\psi)}{\partial\psi_{l}}\stackrel{\text{!}}{=}0.
\label{psichi2}
 \end{equation}
In Eq.~\ref{psichi2}, $l$ runs over all grid pixels. 
The weak lensing term in Eq.~\ref{overall_chi} is derived from Eq.~\ref{epsilon} with a measured average complex ellipticity of background sources in each grid pixel $\epsilon$
\begin{equation}
  \chi^{2}_{\mathsf w}=\sum\limits_{i,j}(\varepsilon -g(\psi))_{i}\mathcal{C}^{-1}_{ij}(\varepsilon -g(\psi))_{j}.
  \label{chi2weak}
\end{equation}
The covariance matrix $\mathcal{C}$ is non-diagonal because the algorithm adaptively averages over a number of background-ellipticity measurements ($\sim10$) in each pixel to account for the intrinsic ellipticity of background sources. Due to this averaging scheme, neighboring pixels may share a certain number of background sources and the algorithm keeps track of these correlations between pixels as described in \citet{Merten2009}. The connection to the lensing potential is given by Eq.~\ref{reducedshear} which, when inserted into Eq.~\ref{chi2weak}, yields
\begin{equation}
 \chi^{2}_{\mathsf w}(\psi)=\sum\limits_{i,j}\left(\varepsilon-\frac{Z(z)\gamma(\psi)}{1-Z(z)\kappa(\psi)}\right)_{i}\mathcal{C}^{-1}_{ij}\left(\varepsilon-\frac{Z(z)\gamma(\psi)}{1-Z(z)\kappa(\psi)}\right)_{j},
\label{chi2weak_2}
\end{equation}
where again both indices $i$ and $j$ run over all grid cells. Note that all lensing quantities given by Eq.~\ref{lensingquantities} have a redshift dependence introduced by the critical density in Eqs.~\ref{lensingpotential},~\ref{critical_dens}. This is taken into account by a cosmological weight function \citep{Bartelmann2001} scaling each pixel to a fiducial redshift of infinity during the reconstruction.  
\begin{equation}
  Z(z):=\frac{D_{\infty}D_{\mathsf ls}}{D_{\mathsf l\infty}D_{\mathsf s}}H(z-z_{\mathsf l})
\label{cosmoweight}
\end{equation}
The Heaviside step function ensures that only sources behind the lens redshift $z_{\mathrm{l}}$ have non-zero weight. 

The definition of the strong lensing term in Eq.~\ref{overall_chi} makes use of the fact the position of the lens' critical line at a certain redshift can be inferred from the position of multiple images. It has been shown in \citet{Merten2009} and \citet{Meneghetti2010a} that pixel sizes $> 5\arcsec$ are large enough to make this simple assumption. Therefore, following Eq.~\ref{Jacobiandeterminant}
\begin{equation}
 \chi^{2}_{\mathsf s}(\psi)=\frac{|\det\mathcal{A}(\psi)|^{2}_{i}}{\sigma^{2}_{i,\mathsf{s}}}=\frac{|(1-Z(z)\kappa(\psi))^{2}-|Z(z)\gamma(\psi)|^{2}|^{2}_{i}}{\sigma^{2}_{i,\mathsf{s}}},
\label{chi2strong}
\end{equation}
where this term is only assigned to those grid cells which are part of the critical line at a certain redshift $z$ given the positions of multiple images. 
The error term $\sigma$ is then given by the cell size of the grid following
\begin{equation}
 \sigma_{\mathsf s}\approx \left.\frac{\partial\det\mathcal{A}}{\partial \theta}\right|_{\theta_{\mathsf c}}\delta\theta\approx \frac{\delta\theta}{\theta_{\mathsf E}},
\label{chi2strong_2}
\end{equation}
with $\theta_{\mathsf E}$ being an estimate of the Einstein radius of the lens. 

The missing connection to the lensing potential $\psi$ is given by Eq.~\ref{lensingquantities}. The numerical technique of finite differencing is then used to express the basic lensing quantities by simple matrix multiplications
\begin{align}
 \kappa_{i}&=\mathcal{K}_{ij}\psi_{j} 
\label{c} \\
\gamma^{1}_{i}&=\mathcal{G}^{1}_{ij}\psi_{j} 
\label{s1} \\
\gamma^{2}_{i}&=\mathcal{G}^{2}_{ij}\psi_{j}
\label{s2}
\end{align}
where $\mathcal{K}_{ij}, \mathcal{G}^{1}_{ij}$ and $\mathcal{G}^{2}_{ij}$ are sparse matrices representing the finite differencing stamp of the respective differential operator \citet{Seitz1998,Bradavc2005,Merten2009}.
With these identities in mind it can be shown that Eq.~\ref{psichi2} takes the form of a linear system of equations, which is solved numerically.
There are two important aspects to this method, which we will only mention briefly. First, a two-level iteration scheme is employed to deal with the non-linear nature of the reduced shear \citep{Schneider1995} and to avoid overfitting of local noise contributions \citep{Merten2009}. Second, a regularization scheme is adapted \citep{Seitz1998,2000} to ensure a smooth transition from one iteration step to the next. In this work we adapt the regularization scheme of \citet{Bradavc2005}. 

It is important that a sophisticated and numerically involved lensing inversion algorithm is tested thoroughly and under controlled but realistic conditions. These tests were performed in \citet{Meneghetti2010a}, where \texttt{SaWLens} showed its ability to routinely and reliably reconstruct the mass distribution of simulated galaxy clusters at the 10\% accuracy level from small ($\sim$ 50kpc) to large scales (several Mpc). Other methods relying on either strong or weak lensing constraints are limited to either small or large scales and showed a much larger scatter of $\sim$ 20\%.  
Also, the method which we have described in this section has been successfully used in the reconstruction of observed galaxy clusters \citep{Merten2009,Merten2011,Umetsu2012,Medezinski2013,Patel2013}.

\section{The CLASH Data Set}\label{Data}
Our analysis focuses on the X-ray selected sub-sample of CLASH (Table~\ref{CLASH_CLUSTERS}). For each of these clusters a large number of lensing constraints was collected, either from the HST CLASH survey \citep{Postman2012}, the accompanying Subaru/Suprime-Cam \citep{Postman2012, Umetsu2011, Medezinski2013} or ESO/WFI \citep{Gruen2013} weak lensing observations or from the CLASH-VLT spectroscopic program \citep{Balestra2013}. The data collection includes strong-lensing multiply-imaged systems together with accurate spectroscopic or photometric redshifts and weak-lensing shear catalogs on the full cluster field, paired with a reliable background selection of weak lensing sources. 

\subsection{Strong Lensing in the HST Fields}
The \citet{Zitrin2009} method is applied to identify multiple-image systems in each cluster field. 
The respective strong-lensing mass models for several CLASH clusters have already been published \citep{Zitrin2011e,Zitrin2012,Zitrin2012a,Zitrin2013,Coe2012,Coe2013,Umetsu2012,Zheng2012} and the full set of strong-lensing models and multiple-image identifications will be presented in Zitrin et al.~(2014 in prep). 
Exceptions are the cluster RXC~J2248, where the multiple-image identification is based on the \citet{Monna2014} strong-lensing mass model, and RX~J1532, where our team was not able to identify any strong-lensing features to date. In this case, we derive the underlying lensing potential from weak lensing only with a significantly coarser resolution in the central region, compared to the strong-lensing clusters.

\begin{figure}
 \begin{center}
  \includegraphics[width=.48\textwidth]{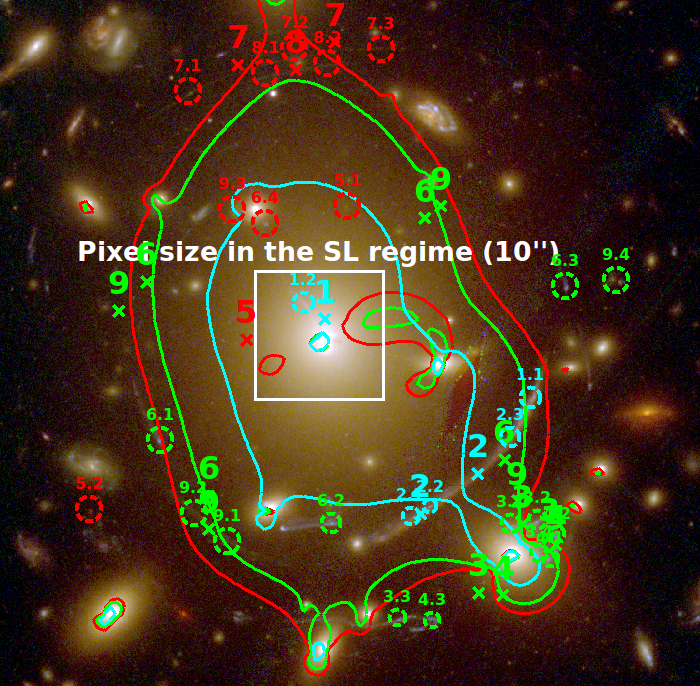}\\
 \end{center}
  \caption{Estimation of the critical line for the \texttt{SaWLens} analysis of Abell~383. Shown by the labeled circles are the different sets of multiple-image systems identified by \citet{Zitrin2011e}. The three
  solid lines show the critical lines from their strong lensing model for three different source redshifts (cyan: $z_{s}=1.01$, green: $z_{s}=2.55$ and red: $z_{s}=6.03$). The crosses with integer labels
  show our critical line estimate for a particular multiple image system with the same ID number. The white box shows the \texttt{SaWLens} pixel size in the strong-lensing regime. The critical line estimates and the multiple-image systems
  are divided into three groups. Cyan indicates systems at $z_{s}=1.01$, green contains systems in a redshift range between $z_{s}=2.20$ and $z_{s}=3.90$ and red systems in the range
  from $z_{s}=4.55$ to $z_{s}=6.03$  
  \label{Crit_line}}
\end{figure}

A summary of multiple-image systems found in each cluster is given in Table~\ref{HSTSL_Sample_Table}. From the identified multiple images we estimate the locations of critical lines following the approach of \citet{Merten2009}.
We show this critical line estimation for one concrete example in Fig.~\ref{Crit_line}, where we indicate the multiple images identified by \citet{Zitrin2011e} in Abell~383 together with the critical lines derived from a detailed strong-lensing
model of the cluster. In addition we show our critical line estimation from the multiple-image identifications which is in excellent agreement with the critical lines from the strong-lensing model given the pixel size of our reconstruction. 

Redshifts for all strong lensing features are either taken from the literature, spectroscopic redshifts from the on-going CLASH VLT-Vimos large program (186.A-0798) \citep{Balestra2013} or from the CLASH photometry directly using Bayesian photometric redshifts \citep[BPZ,][]{Benitez2000}. CLASH has been explicitly designed to deliver accurate photometric redshifts for strong lensing features \citep{Postman2012}. The accuracy of the CLASH photometric redshifts has been recently evaluated in \citet{Jouvel2014} where we found 3.0\%(1+z) precision for strong-lensing arcs and field galaxies. 

\begin{deluxetable*}{lccccc}
 \tablecaption{Strong-lensing constraints\label{HSTSL_Sample_Table}}
 \tablewidth{0pt}
 \tablehead{
   \colhead{Name} & \colhead{$N_{\textrm{sys}}$\tablenotemark{a}} &\colhead{$N_{\textrm{spec}}$\tablenotemark{b}}& \colhead{$N_{\textrm{crit}}$\tablenotemark{c}}&\colhead{z-range}&\colhead{$<d_{\mathrm{crit}}>$\tablenotemark{d}}\\ 
   &&&&&\colhead{[$\arcsec$]}
   }
   \startdata
Abell~383&	9&	5&	20\tablenotemark{e}&	1.01--6.03&	$17.5\pm	5.7$\\
Abell~209&	6&	0&	5&	1.88--3.5&	$8.5\pm		0.8$\\
Abell~1423&	1&	0&	1&	3.5	&	$17.5\pm$	---\\
Abell~2261&	12&	0&	18&	1.54--4.92&	$18.1\pm	8.2$\\
RXJ2129+0005&	4&	1&	8&	0.55--1.965&	$8.1\pm		3.5$\\
Abell 611&	4&	3&	9&	0.908--2.59&	$13.1\pm	4.5$\\
MS2137-2353&	2&	2&	6&	1.501--1.502&	$12.2\pm	4.7$\\
RXCJ2248-4431&	11&	10&	22&	1.0--6.0&	$27.8\pm	5.6$\\
MACSJ1115+0129&	2&	0&	5&	2.46--2.64&	$19.9\pm	9.2$\\
MACSJ1931-26&	7&	0&	8&	2.6--3.95&	$29.2\pm	1.3$\\
RXJ1532.8+3021&	0&	0&	0&	\nodata&		\nodata\\
MACSJ1720+3536&	7&	0&	11&	0.6--4.6&	$19.3\pm	8.8$\\
MACSJ0429-02&	3&	0&	6&	1.6--4.1&	$11.8\pm	3.6$\\
MACSJ1206-08&	13&	4&	33&	1.033--5.44&	$28.1\pm	14.8$\\
MACSJ0329-02&	6&	0&	12&	1.55--6.18&	$23.7\pm	5.2$\\
RXJ1347-1145&	13&	1&	15&	0.7--4.27&	$31.6\pm	13.3$\\
MACSJ1311-03&	2&	0&	4&	2.63--6.0&	$12.9\pm	5.3$\\
MACSJ1423+24&	5&	3&	18&	1.779--2.84&	$15.0\pm	5.6$\\
MACSJ0744+39&	5&	0&	8&	1.15--4.62&	$31.6\pm	16.2$\\
CLJ1226+3332&	4&	0&	9&	2.0--4.2&	$23.2\pm	12.2$\\
   \enddata
\tablenotetext{a}{The number of multiple-image systems in this cluster field.}
\tablenotetext{b}{The number of spectroscopically confirmed multiple-image systems.}
\tablenotetext{c}{The number of critical line estimators derived from the position of multiple-image systems.}
\tablenotetext{d}{The mean distance and its standard deviation from the cluster center to the critical line estimators.}
\tablenotetext{e}{An illustration of how the critical line estimators for this specific systems were derived is given in Fig.~\ref{Crit_line}}
\end{deluxetable*}

\subsection{Weak Lensing in the HST Fields}
\label{WL_HST}
\begin{figure*}
 \begin{center}
 \includegraphics[width=.95\textwidth]{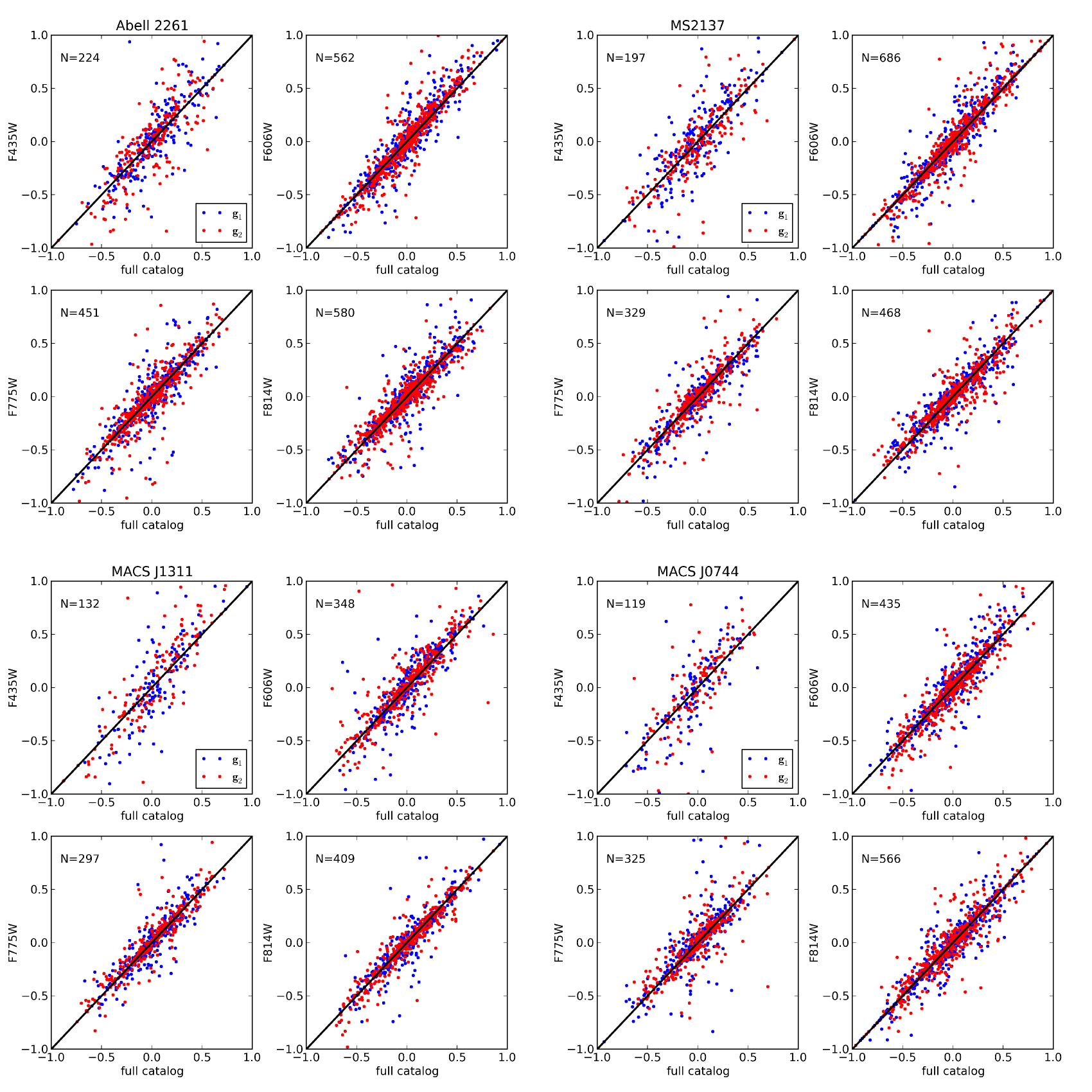}
 \end{center}
 \caption{The correlation of shape measurements in different HST/ACS filters. This analysis is carried out for four different 
 clusters. A massive low redshift cluster (A2261), a known relaxed system at intermediate redshift (MS2137), a system at intermediate 
 redshift (M1311) and a system of known complex morphology at relatively high redshift \citep[MACS~J0744,][]{Korngut2011}. The \textit{upper left} panel 
 of each group shows the correlation of ellipticities measured in the F435W images compared to the combined HST/ACS catalogs.
 The \textit{upper right}, \textit{lower left} and \textit{lower right panels} show the same correlation for the F606W, F775W and F814W catalogs. 
 Also shown in each individual plot is the number of overlapping galaxies in the different catalogs. 
 \label{HST_correlations}}
\end{figure*}

For cluster mass reconstruction, the HST delivers a four to five times higher density of weakly lensed background galaxies than observations from the ground \citep[e.g.][]{Clowe2006,Bradav2006,Bradav2008,Merten2011,Jee2012}. We use a modified version of the MosaicDrizzle pipeline \citep{Koekemoer2002,Koekemoer2011} for CLASH HST imaging. In each filter, shape analyses are not performed on the full image stack but on each visit and in each filter separately. This is necessary in order not to mix different  point-spread-functions (PSF) present across the field due to the time separation (several days) of the different visit exposures. All individual images were combined 
using a drizzle pixel scale of 0.03 \arcsec. 

All CLASH ACS images are corrected for charge-transfer-inefficiency (CTI) \citep[e.g.][]{Massey2010a,Anderson2010}. For shape measurement and PSF correction we use the RRG package \citep{Rhodes2000}, which implements an HST breathing model \citep{Leauthaud2007, Rhodes2007a} to correct for the thermally induced variation of the HST PSF. The method has been used for cosmic shear \citep{Massey2007} and cluster lensing \citep{Bradav2008,Merten2011} applications. The level of PSF variation was determined from the inspection of stars in the field of each visit \citep{Rhodes2007a} and by cross-comparison with the STScI focus tool\footnote[1]{\href{http://www.stsci.edu/hst/observatory/focus}{http://www.stsci.edu/hst/observatory/focus}}. The implementation and accuracy of this tool clearly complies with the needs of cluster lensing 
applications \citep[][and references therein]{2008}. 
Shape measurement and PSF corrections are performed for each individual ACS exposure in each filter. 
Every shear catalog is then rotated into a North-up orientation in order to have
a unique orientation reference for the directional shape parameters and individual visits are 
combined using a signal-to-noise weighted average for multiple measurements of the same object.
Catalogs in different filters are combined by using a signal-to-noise weighted average for matching sources. For all but two clusters, seven broadband ACS filters (F435W, F475W, F606W, F625W, F775W, F814W and F850L) were used for shape analysis from CLASH and archival observations \citep{Postman2012}. In the case of Abell~611 we did not use F606W and F625W images since the focus tool did not cover the time period when these observations were taken. In the case of RX~J2129 additional F555W data is included from archival data. 

The lensed background sample for each combined catalog was selected using two photo-z criteria. First, the most likely redshift according to the probability distribution of BPZ had to be at least 20\% larger than the cluster redshift to ensure a limited contamination by cluster members. Second, the lower bound on the source redshift (based on the BPZ probability distribution) had to be larger or equal to the cluster redshift. A size cut and removal of obvious artifacts finalizes each HST weak lensing catalog and the effective lensing redshift of the background distribution is determined from the photometric redshift of each object in the final catalog.
All relevant information about the HST weak lensing catalogs is summarized in Table~\ref{HSTWL_Sample_Table}. 

The cross-shear component was found to be small at all radii. To see this in the case of our HST weak lensing we refer to the panels for Abell~1423 and CL~J1226 in Fig.~\ref{shear_profiles}. We also found strong correlations in both ellipticity components between different ACS filter measurements. This is demonstrated for four different filters and four different clusters in Fig.~\ref{HST_correlations}. As a final cross-check we performed lensing inversions of the HST weak lensing data only, as it is shown for the case of Abell~1423 in Fig.~\ref{convergence_maps}; all of these showed strong 
correlations with the light distribution in the HST fields. All our systematic tests will be presented in greater detail in a focused CLASH HST weak-lensing analysis by Melchior et al.~(2014 in prep.). 
\begin{deluxetable*}{lcccccc}
 \tablecaption{HST weak-lensing constraints\label{HSTWL_Sample_Table}}
 \tablewidth{0pt}
\tablehead{
   \colhead{Name} &\colhead{$N_{\mathrm{band}}$\tablenotemark{a}}&\colhead{$N_{\mathrm{gal}}$\tablenotemark{b}}&\colhead{$\rho_{\mathrm{gal}}$\tablenotemark{c}}&\colhead{$z_{\textrm{eff}}$}\tablenotemark{d}&\colhead{$<\epsilon_{1}>$\tablenotemark{e}}&\colhead{$<\epsilon_{2}>$\tablenotemark{f}}\\
   &&&\colhead{[$\textrm{arcmin}^{-2}$]}&&&
   }
   \startdata
Abell~383&		7&	796&	50.7&	0.90&	$0.003\pm0.30$&		$0.023\pm0.29$\\
Abell~209&		7&	832&	44.0&	0.95&	$0.039\pm0.29$&		$-0.016\pm0.28$\\
Abell~1423&		7&	807&	50.3&	0.92&	$0.009\pm0.29$&		$0.007\pm0.28$\\
Abell~2261&		7&	725&	46.7&	0.79&	$0.012\pm0.28$&		$-0.008\pm0.29$\\
RXJ2129+0005&		8&	624&	35.8&	0.82&	$-0.025\pm0.30$&	$0.012\pm0.30$\\
Abell 611&		5&	547&	42.3&	0.86&	$-0.001\pm0.27$&	$0.021\pm0.29$\\
MS2137-2353&		7&	801&	48.3&	1.12&	$0.011\pm0.32$&		$0.014\pm0.33$\\
RXCJ2248-4431&		7&	598&	38.5&	1.12&	$-0.004\pm0.28$&	$0.039\pm0.29$\\
MACSJ1115+0129&		7&	491&	37.4&	1.03&	$-0.007\pm0.30$&	$0.004\pm0.30$\\
MACSJ1931-26&		7&	709&	59.5&	0.82&	$-0.035\pm0.23$&	$0.013\pm0.24$\\
RXJ1532.8+3021&		7&	508&	35.9&	1.07&	$0.018\pm0.29$&		$0.020\pm0.29$\\
MACSJ1720+3536&		7&	635&	40.6&	1.11&	$0.035\pm0.30$&		$-0.005\pm0.30$\\
MACSJ0429-02&		7&	654&	42.4&	1.08&	$0.003\pm0.29$&		$-0.001\pm0.29$\\
MACSJ1206-08&		7&	581&	51.2&	1.13&	$-0.007\pm0.31$&	$-0.005\pm0.28$\\
MACSJ0329-02&		7&	493&	35.2&	1.18&	$-0.004\pm0.29$&	$-0.038\pm0.31$\\
RXJ1347-1145&		7&	633&	45.7&	1.13&	$-0.004\pm0.32$&	$0.023\pm0.31$\\
MACSJ1311-03&		7&	447&	33.7&	1.03&	$0.020\pm0.27$&		$0.003\pm0.32$\\
MACSJ1423+24&		7&	899&	75.3&	1.04&	$0.016\pm0.31$&		$0.042\pm0.31$\\
MACSJ0744+39&		7&	743&	61.3&	1.32&	$-0.041\pm0.30$&	$0.001\pm0.31$\\
CLJ1226+3332&		7&	925&	32.7&	1.66&	$-0.010\pm0.33$&	$0.002\pm0.33$\\
\enddata
\tablenotetext{a}{The number of HST/ACS bands from which the final shear catalog was created.}
\tablenotetext{b}{The number of background selected galaxies in the shear catalog.}
\tablenotetext{c}{The surface-number density of background selected galaxies in the field.} 
\tablenotetext{d}{The effective redshift of the background sample, derived from their photo-zs and by calculating the average of the $D_{ls}/D_{s}$ ratio and correcting for the non-linearity of the reduced shear.} 
\tablenotetext{e}{The mean value and standard deviation of the first component of the ellipticity in the total field.}
\tablenotetext{f}{As above but for the second component.}
\end{deluxetable*}

\begin{deluxetable*}{lcccccc}
 \tablecaption{Ground-based weak-lensing constraints\label{SubaruWL_Sample_Table}}
 \tablewidth{0pt}
 \tablehead{
   \colhead{Name} &\colhead{shape-band}&\colhead{$N_{\mathrm{gal}}$}&\colhead{$\rho_{\mathrm{gal}}$}&\colhead{$z_{\textrm{eff}}$}&\colhead{$<\epsilon_{1}>$}&\colhead{$<\epsilon_{2}>$}\\
   &&&\colhead{[$\textrm{arcmin}^{-2}$]}&&&
   }
   \startdata
Abell~383&	Ip&	7062&	9.0&	1.16&	$-0.014\pm0.35$&	$-0.001\pm0.36$\\
Abell~209&	Rc&	14694&	16.4&	0.94&	$-0.009\pm0.35$&	$0.001\pm0.35$\\
Abell~1423\tablenotemark{a}&	\nodata&	\nodata&	\nodata&	\nodata&	\nodata&			\nodata\\
Abell~2261&	Rc&	15429&	18.8&	0.89&	$0.002\pm0.33$&		$0.004\pm0.32$\\
RXJ2129+0005&	Rc&	20104&	24.5&	1.16&	$-0.006\pm0.35$&	$0.001\pm0.35$\\
Abell 611&	Rc&	7872&	8.8&	1.13&	$0.013\pm0.41$&		$-0.001\pm0.41$\\
MS2137-2353&	Rc&	9133&	11.6&	1.23&	$0.003\pm0.36$&		$-0.006\pm0.35$\\
RXCJ2248-4431&	WFI~R&	4008&	5.5&	1.05&	$-0.005\pm0.32$&	$0.011\pm0.33$\\
MACSJ1115+0129&	Rc&	13621&	15.1&	1.15&	$-0.008\pm0.39$&	$-0.007\pm0.39$\\
MACSJ1931-26&	Rc&	4343&	5.3&	0.93&	$0.009\pm0.56$&		$0.005\pm0.53$\\
RXJ1532.8+3021&	Rc&	13270&	16.6&	1.15&	$-0.001\pm0.37$&	$-0.002\pm0.37$\\
MACSJ1720+3536&	Rc&	9855&	12.5&	1.13&	$0.010\pm0.37$&		$-0.020\pm0.38$\\
MACSJ0429-02&	Rc&	9990&	12.0&	1.25&	$0.002\pm0.41$&		$0.001\pm0.41$\\
MACSJ1206-08&	Ic&	12719&	13.7&	1.13&	$-0.001\pm0.37$&	$0.003\pm0.37$\\
MACSJ0329-02&	Rc&	25427&	29.5&	1.18&	$-0.002\pm0.32$&	$-0.001\pm0.32$\\
RXJ1347-1145&	Rc&	9385&	8.9&	1.17&	$0.001\pm0.53$&		$-0.002\pm0.53$\\
MACSJ1311-03&	Rc&	13748&	20.2&	1.07&	$0.004\pm0.40$&		$0.008\pm0.41$\\
MACSJ1423+24&	Rc&	7470&	9.8&	0.98&	$-0.001\pm0.42$&	$-0.003\pm0.41$\\
MACSJ0744+39&	Rc&	7561&	9.5&	1.41&	$-0.008\pm0.39$&	$-0.010\pm0.39$\\
CLJ1226+3332\tablenotemark{a}&	\nodata&	\nodata&	\nodata&	\nodata&	\nodata&			\nodata\\	
\enddata
\tablenotetext{a}{No ground-based data of sufficient data quality in terms of seeing, exposure time and band coverage was available at the time this work was published.}
\tablecomments{These values derive from the comprehensive CLASH weak lensing work by \citet{Umetsu2014}. Column explanations are identical to Table~\ref{HSTWL_Sample_Table}.}
\end{deluxetable*}

\subsection{Weak Lensing in the Ground-Based Fields}
\label{WL_Subaru}
The creation of our weak lensing shear catalogs from ground-based data is described in Sec.~4 of \citet{Umetsu2014}. 
For completeness we summarize the properties of these catalogs in Table~\ref{SubaruWL_Sample_Table}.

\subsection{Combination of Shear Catalogs}
\begin{figure*}
 \begin{center}
 \includegraphics[width=.95\textwidth]{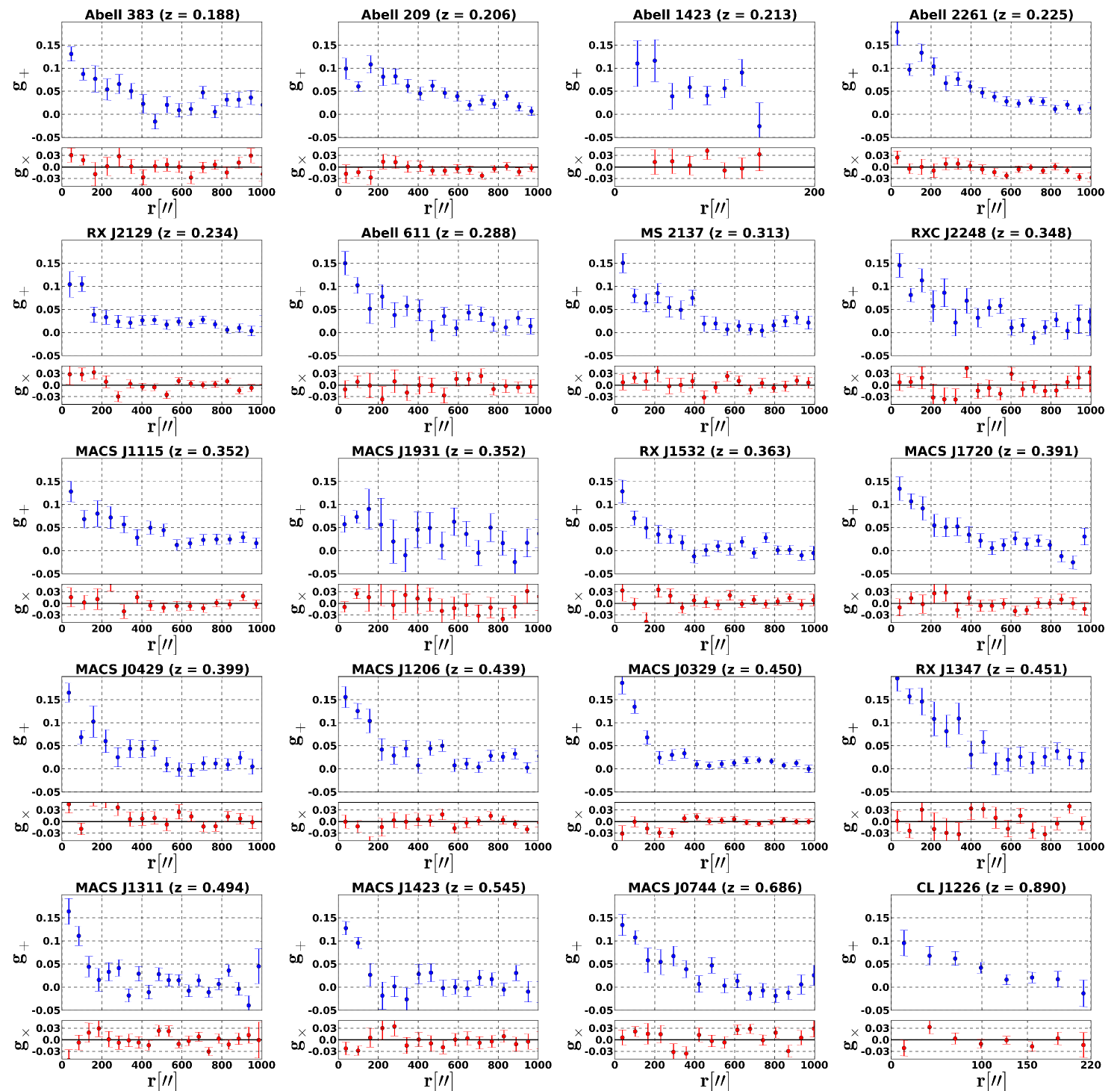}
 \end{center}
 \caption{Shear profiles for the final ellipticity catalogs of 20 X-ray selected CLASH clusters. In the case of Abell~1423 and CL~J1226
 these catalogs derive from HST/ACS images only. All other cases show combined HST/ACS and Subaru catalogs. The \textit{top plot} of each panel
 shows the tangential shear profile, the \textit{bottom plot} the cross shear profile with respect to the cluster center defined in Table~\ref{CLASH_CLUSTERS}.
 1-$\sigma$ error bars were derived from 250 bootstrap resamplings of each input catalog.
 \label{shear_profiles}}
\end{figure*}

We combine the HST and ground-based catalogs into a single weak lensing catalog before the \texttt{SaWLens} reconstruction. In order to do so, 
we first correct for the different redshifts of the background populations in each catalog. We scale the two shear values in the HST
catalogs with a factor 
\begin{equation}
 \beta=\frac{D_{\mathrm{lS}}D_{\mathrm{H}}}{D_{\mathrm{S}}D_{\mathrm{lH}}},
\end{equation}
which accounts for  the dependence of the shear on the lensing geometry. 
Here, $D_{\mathrm{lS}}$ ($D_{\mathrm{lH}}$) is the angular diameter distance between the lens and the ground-based (HST) sample and $D_{\mathrm{S}}$ ($D_{\mathrm{H}}$) is the angular diameter distance between the observer and the ground-based (HST) sample. 
After applying the correction $\beta$ to the HST shapes, we match the two catalogs by calculating the signal-to-noise weighted 
mean of sources which appear in both catalogs and by concatenating non-matching entries in the two catalogs. 
As a final cross-check we calculate the tangential ($g_{+}$) -and cross-shear ($g_{\times}$) components in azimuthal bins around
the cluster center, which we show in Fig.~\ref{shear_profiles}.

\section{Density Profiles of CLASH Clusters}\label{Results}
Our mass reconstructions with associated error bars are used to fit NFW profiles to the 
surface-mass density distribution. 
Mass and concentration parameters for each of the X-ray selected CLASH clusters are the main result of our observational efforts.
\subsection{Final \texttt{SaWLens} Input and Results}
We summarize the basic parameters of each cluster reconstruction in Table~\ref{Reconstruction_Properties}, including input data, reconstructed field sizes and the refinement levels of the multi-scale grid. 
For two sample clusters, Abell~1423 and CL~J1226, no multi-band wide-field weak lensing data with acceptable seeing and exposure time levels is available. In the case of CL~J1226 this is less severe since 
we have access to a rather wide HST/ACS mosaic and, since the cluster resides at high redshift, the angular size of the reconstruction refers to a large physical size of the system. We therefore include
CL~J1226 in our following mass-concentration analysis, while we drop Abell~1423 from this sample.

The output of the reconstruction is the lensing potential on a multi-scale grid, which is then converted into a convergence or surface-mass density map via Eq.~\ref{lensingquantities}. 
The convergence maps on a wide field for all sample clusters are shown in Fig.~\ref{convergence_maps}. We base our follow-up analysis
on these maps, together with a comprehensive assessment of their error budget. 

\begin{deluxetable*}{lccccccc}
 \tablecaption{Reconstruction properties\label{Reconstruction_Properties}}	
\tablewidth{0pt}
\tablehead{
   \colhead{Name} & \colhead{input data\tablenotemark{a}} &\colhead{Ground-based FOV} &\colhead{HST FOV}&\colhead{$\Delta_{\textrm{ground}}$\tablenotemark{b}}&\colhead{$\Delta_{\textrm{ACS}}$\tablenotemark{c}}&\colhead{$\Delta_{\textrm{SL}}$\tablenotemark{d}}&\colhead{$\#_{masks}$\tablenotemark{e}}\\
   &&\colhead{[$\arcsec\times\arcsec$]}&\colhead{[$\arcsec\times\arcsec$]}&\colhead{[$\arcsec$]}&\colhead{[$\arcsec$]}&\colhead{[$\arcsec$]}&\\
   }
   \startdata
Abell~383&	S,~A,~H&	1500$\times$1500&	173$\times$173& 29&	12&	10&	2\\
Abell~209&	S,~A,~H&	1500$\times$1500&	150$\times$150&	25&	12&	8&	2\\
Abell~1423&	H&		\nodata&		200$\times$200&	\nodata&	13&	\nodata&	\nodata\\	
Abell~2261&	S,~A,~H&	1500$\times$1500&	150$\times$150&	25&	13&	8&	2\\
RXJ2129+0005&	S,~A,~H&	1500$\times$1500&	150$\times$150&	25&	10&	8&	3\\
Abell 611&	S,~A,~H&	1400$\times$1400&	168$\times$168&	28&	10&	9&	1\\
MS2137-2353&	S,~A,~H&	1500$\times$1500&	180$\times$180&	30&	14&	10&	1\\
RXCJ2248-4431&	W,~A,~H&	1500$\times$1500&	171$\times$171&	34&	12&	11&	7\\
MACSJ1115+0129&	S,~A,~H&	1500$\times$1500&	150$\times$150&	25&	10&	8&	2\\
MACSJ1931-26&	S,~A,~H&	1500$\times$1500&	179$\times$179&	36&	10&	10&	0\\
RXJ1532.8+3021&	S,~A   &	1500$\times$1500&	155$\times$155&	26&	10&	\nodata&	3\\
MACSJ1720+3536&	S,~A,~H&	1500$\times$1500&	150$\times$150&	25&	9&	8&	3\\
MACSJ0429-02&	S,~A,~H&	1500$\times$1500&	167$\times$167&	28&	10&	9&	3\\
MACSJ1206-08&	S,~A,~H&	1500$\times$1500&	150$\times$150&	25&	12&	8&	0\\
MACSJ0329-02&	S,~A,~H&	1500$\times$1500&	150$\times$150&	25&	9&	8&	0\\
RXJ1347-1145&	S,~A,~H&	1500$\times$1500&	180$\times$180&	30&	12&	10&	1\\
MACSJ1311-03&	S,~A,~H&	1500$\times$1500&	150$\times$150&	25&	10&	8&	6\\
MACSJ1423+24&	S,~A,~H&	1500$\times$1500&	155$\times$155&	26&	8&	8&	2\\
MACSJ0744+39&	S,~A,~H&	1500$\times$1500&	150$\times$150&	30&	9&	7&	4\\
CLJ1226+3332&	A,~H&		\nodata		 &	300$\times$300&	\nodata&	8&	6&	0\\
\enddata
\tablenotetext{a}{'S' stands for Subaru weak lensing data, 'W' stands for ESO/WFI weak lensing data, 'A' stands for HST/ACS weak lensing data and 'H' stands for HST strong lensing data.}
\tablenotetext{b}{The pixel size of the grid in the Subaru or ESO/WFI weak lensing regime.}
\tablenotetext{c}{The pixel size of the grid in the HST/ACS weak lensing regime.}
\tablenotetext{d}{The pixel size of the grid in the strong lensing regime.}
\tablenotetext{e}{The number of masks in the reconstruction grid. There are necessary if bright stars blend large portions of the FOV.}
\end{deluxetable*}

\subsection{Error Estimation}
Non-parametric methods, especially when they include non-linear constraints in the strong-lensing regime, do not offer a straight-forward way to analytically describe the error bars attached to reconstructed quantities \citep{2000}. We therefore follow 
the route of resampling the input catalogs to obtain error bars on our reconstructed convergence maps. The weak-lensing input is treated by bootstrap resampling the shear catalogs \citep[see e.g.][]{Bradavc2005,Merten2011}. 
For the strong-lensing input, we randomly sample the allowed redshift range for each multiple-image system and randomly discard systems from the strong-lensing sample which were identified only as candidate systems by the \citet{Zitrin2009} method.
We sequentially repeat the reconstructions using 1000 resampled realizations for each cluster reconstruction. This number  
is chosen somewhat arbitrarily but is primarily driven by runtime considerations, due to the high numerical demands of non-parametric reconstruction methods. 
From the observed scatter in the ensemble of realizations we derive our error bars, e.g. in the form of a covariance matrix for binned convergence profiles, as we describe them in the following section. 

\subsection{From Convergence Maps to NFW Profile Parameters}
\begin{figure}
 \begin{center}
  \includegraphics[width=.42\textwidth]{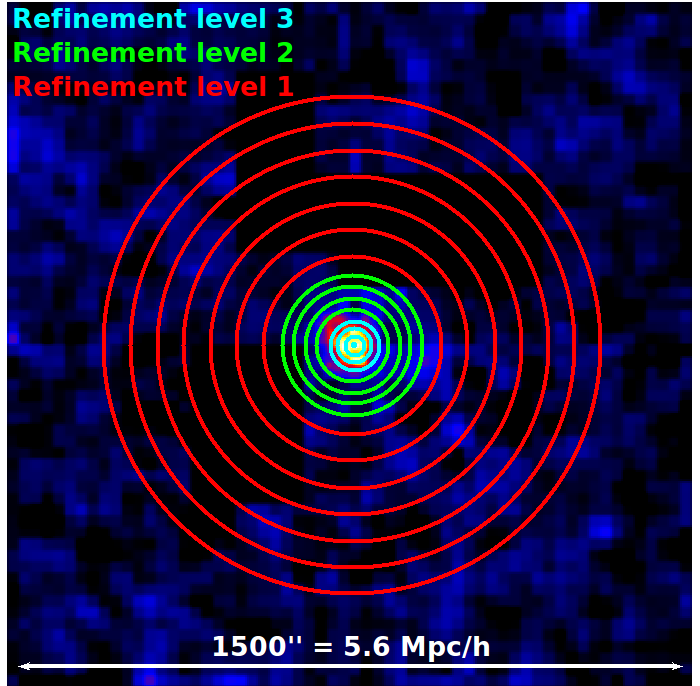}\\
 \end{center}
  \caption{The adaptive binning scheme for the radial convergence profiles. Shown in this figure are the actual bins, overlaid on the cluster's  convergence map, used to derive
  the convergence profile for the cluster MACS J1720 (compare Fig.~\ref{convergence_profiles}). The size of the bins follows
  the three levels of grid refinement as they are visualized in Fig.~\ref{AdaptGrid} and listed in Table~\ref{Reconstruction_Properties}. 
  \label{binnin_scheme}}
\end{figure}

Additional steps are needed to go from non-parametric maps of the surface-mass density distribution to an actual NFW fit of the halo. 
First, since we are interested in 1D density profiles, we apply an azimuthal binning scheme, with a bin pattern that follows the adaptive
resolution of our multi-scale maps. The initial bin is limited by the resolution of the highest refinement level of the convergence 
grid (compare Table~\ref{Reconstruction_Properties}) and the outer-most bin is set to a physical scale of 2 Mpc/h for each halo.
We split the radial range defined by the two thresholds
into 15 bins. An example for the cluster MACS~J1720 is shown in Fig.~\ref{binnin_scheme}. An exception is 
CL~J1226 with no available wide-field data from the ground, where we were limited to a maximum radius of 1.2 Mpc/h and where we divided the radial
range into 11 bins. The center for the radial profile is the dominant peak in the convergence map.
We applied this binning scheme to all 1000 convergence realizations for each cluster and derived the covariance
matrix for the convergence bins. Both the convergence data points and the convergence matrix are shown in Fig.~\ref{convergence_profiles}.

To the convergence bins and the corresponding covariance matrix we fit a NFW profile given by Eq.~\ref{NFW}.
We numerically project the NFW profile on a sphere along the
line-of-sight and thereby introduce the assumption of spherical symmetry in our cluster mass profiles. This is certainly not justified
for all sample clusters and may introduce biases. We discuss this issue in further detail in Sec.~\ref{SMc}.

\begin{figure}
 \begin{center}
  \includegraphics[width=.45\textwidth]{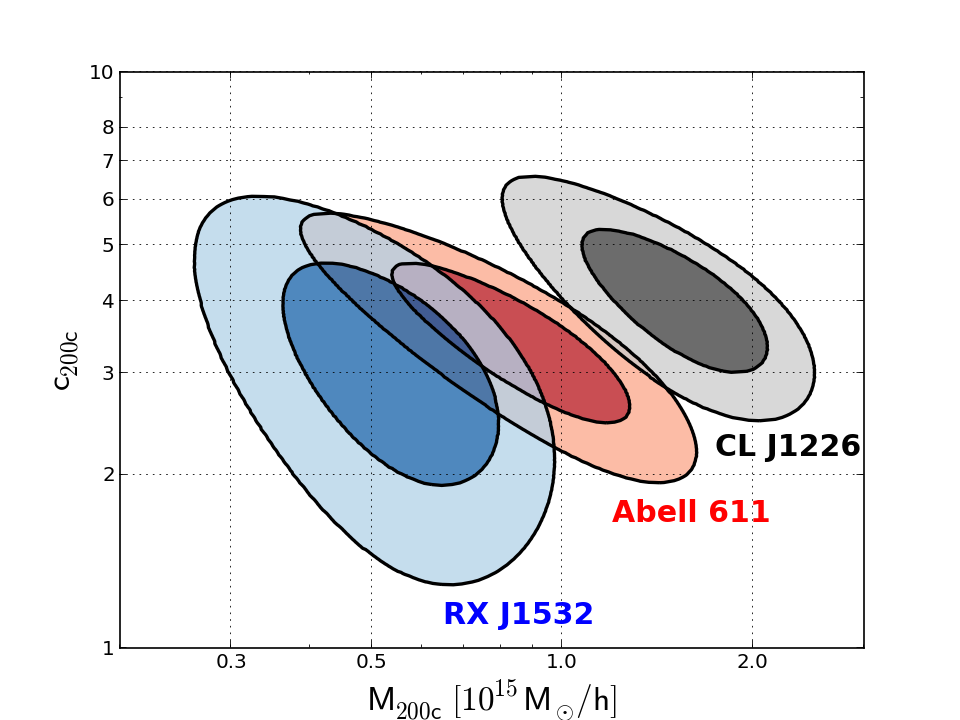}
 \end{center}
  \caption{The likelihood of NFW fits in the c-M plane. The cluster Abell~611 represents a typical CLASH cluster at an intermediate
  redshift with the full set of lensing constraints available. RX~J1532 is the only cluster in this c-M analysis without strong
  lensing constraints and CL~J1226 is the only cluster in the sample without available Subaru weak lensing data. The inner and outer
  contours show the 68\% and 95\% confidence levels.
  \label{c_M_likelihood}}
\end{figure}

We perform the profile fitting using the least-squares formalism by minimizing
\begin{equation}
\chi^{2}(\vec{p})=\sum\limits_{i,j=0}^{N_{\mathrm{bin}}}\left(\kappa_{\mathrm{bin}}-\kappa(\vec{p})\right)_{i}\mathcal{C}^{-1}_{ij}\left(\kappa_{\mathrm{bin}}-\kappa(\vec{p})\right)_{j},
\end{equation}
where $\vec{p}=(\rho_{\mathrm{s}},r_{\mathrm{s}})$ and $\mathcal{C}$ is the covariance matrix of the binned data. The numerical
fitting is performed using the open-source library \texttt{levmar}\footnote[1]{\href{http://users.ics.forth.gr/~lourakis/levmar/}{http://users.ics.forth.gr/$\sim$lourakis/levmar/}} and by making use
of the Cholesky decomposition of $\mathcal{C}^{-1}$. The best-fit parameters, the corresponding covariance matrix and the fitting norm is reported in Table~\ref{NFW_FIT_Table}.
We use these values, together with Eqs.~\ref{NFW_mass},\ref{concentration} to find our final mass and concentration values
at several different radii. We report this central result of our work 
in Table~\ref{NFW_Mc_Table}.
To visualize degeneracies and to show the information gain when including strong-lensing features into the reconstruction we explore
the likelihood in the concentration-mass plane for three CLASH clusters in Fig.~\ref{c_M_likelihood}.

\subsection{Sources of Systematic Error and Comparison to other Analyses}
Before moving on in our analysis we want to discuss 
possible sources of systematic error. 
In the strong-lensing regime there is the possibility of false identification
of multiple-image systems. In the case of CLASH, many strong-lensing features have no spectroscopic confirmation. 
However, CLASH can rely on 16-band HST photometry for photo-z determination.  
Finally, those systems which are only identified as candidates by the 
\citet{Zitrin2009} method for image identification are considered as such in our extensive bootstrap approach.
Another problem for strong lensing is the shift of multiple-image positions by contributions of projected large scale structure. 
This has been pointed out recently in \citet{DAloisio2011,Host2012}. However,
as it was shown by the latter authors, the expected shift in image postion is well below our minimum reconstruction pixel scale of
5\arcsec~to~10\arcsec~for the different clusters (compare Fig.~\ref{Crit_line}). 

We address shape scatter in the weak-lensing catalogs with the adaptive-averaging approach of the \texttt{SawLens} method
and by bootstrapping the weak-lensing input. 
Foreground contamination of the shear catalogs is another
serious concern which will lead to a significant dilution of the weak lensing signal. In the HST images this is controlled by
reliable photometric redshifts. However, there is the possibility of remaining contamination by cluster members 
in the crowded fields and by stars falsely identified as galaxies. 
Background selection in the ground-based catalogs is more difficult due to the smaller number of photometric bands. 
Hence, we use the \citet{Medezinski2010} method of background selection in color-color space which was optimized
to avoid weak lensing dilution. 

We have not commented yet on the dominant density peak in the lensing reconstruction as our 
center choice for the radial density profile. Due to the inclusion of strong lensing constraints, this peak position 
has an uncertainty of only a few arcsec \citep[e.g.][]{Bradav2006,Merten2011},
but one might argue that e.g. the position of the cluster's BCG is a more accurate tracer of the potential minimum.
However, our pixel resolution is of the order of $\sim 5\arcsec$ and BCG position and the peak in the surface-mass density 
coincide or are offset by one or two pixels. 

More important is the effect of uncorrelated large scale structure \citep[e.g.][and references therein]{Hoekstra2011} 
and tri-axial halo shape \citep{Becker2011} which is picked up by our 
lensing reconstruction. \citet{Becker2011} claim that these effects introduce only small biases in 
the mass determination but increase the scatter by up to 20\% with tri-axial shape being the dominant component.
We do not seek to correct for these effects directly but adapt our way of analyzing numerical simulations accordingly (Sec.~\ref{SMc}).

\begin{figure}
 \begin{center}
  \includegraphics[width=.45\textwidth]{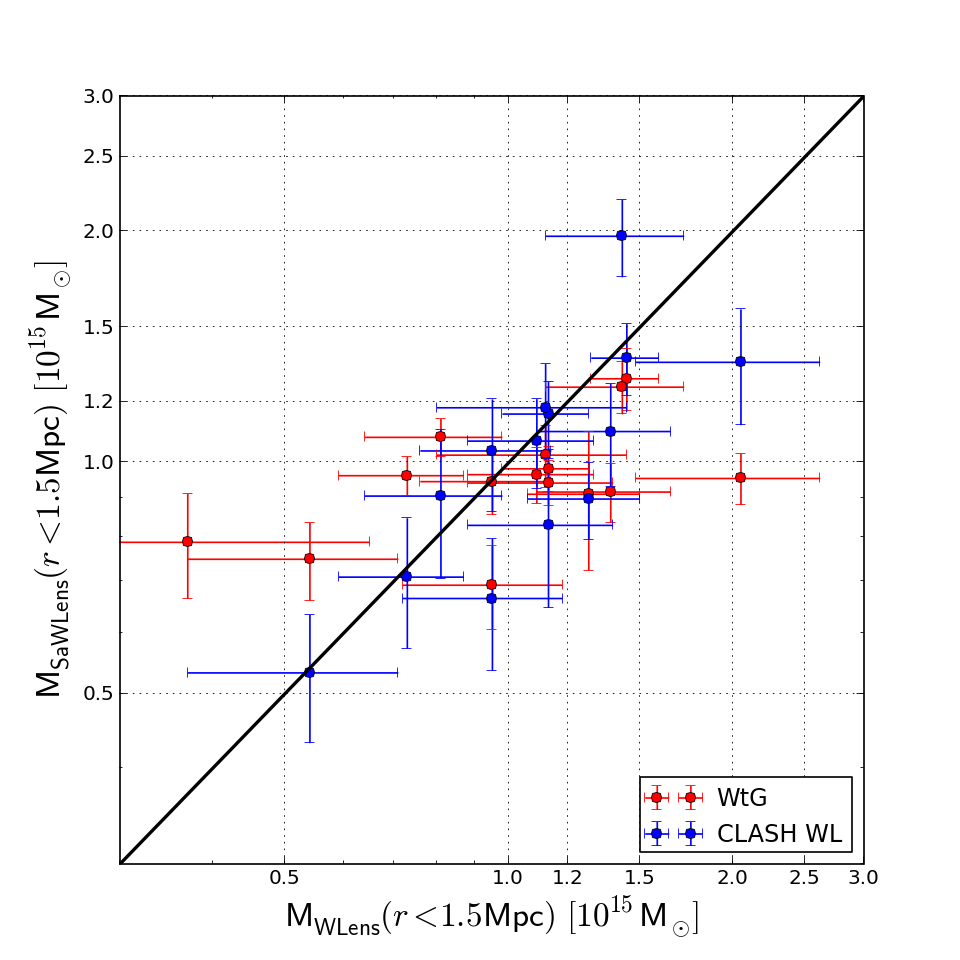}
 \end{center}
  \caption{A comparison between our analysis and other weak-lensing studies. The red data points show clusters in common
  with WtG and the blue data points show the overlap with \citet{Umetsu2014}. On the y-axis we plot enclosed \texttt{SaWLens}
  masses within a radius of 1.5Mpc from the cluster center. The x-axis shows equivalent masses from WtG and U14, respectively. 
  The black line indicates a one-to-one agreement. 
  \label{WtG_comparison}}
\end{figure}

As a final consistency check we look into 15 clusters that we have in common with the Weighing the Giants (WtG) project \citep{2014,Kelly2014,Applegate2014}. 
In addition, we also compare to the CLASH shear and magnification study 
by \citet{Umetsu2014}, which has 14 clusters in common with our analysis. 
For the comparison
we calculate the enclosed mass within 1.5~Mpc of the cluster center following \citet{Applegate2014} 
and show the mass comparison for the three data sets in Fig.~\ref{WtG_comparison}. 
We find median values for the ratios $M_{\textrm{SaWLens}} / M_{\textrm{WtG}}$ and $M_{\textrm{SaWLens}} / M_{\textrm{U14}}$ 
of 0.88 and 0.93, respectively. For the unweighted geometric mean\footnote[1]{The geometric mean satisfies $\left<X/Y\right>=1/\left<Y/X\right>$ for
the ratio of samples $X$ and $Y$.} of these ratios  we find $0.94\pm0.11$ and $0.96\pm0.05$. 
Although we see significant scatter between the different studies,
there is good general agreement. We emphasize that this comparison just serves as a quick cross-check 
between the different mass estimates.
We leave a detailed statistical comparison between different CLASH lensing, X-ray and SZ studies, as well as a comparison to 
earlier results from the literature to a follow-up work by the CLASH collaboration. This will also include a detailed profile
comparison on a cluster-by-cluster basis and as a function of cluster mass and redshift. 

\begin{deluxetable*}{lcccccc}
 \tablecaption{NFW fits: general parameters\label{NFW_FIT_Table}}
\tablewidth{0pt}
\tablehead{
   \colhead{Name} &\colhead{$\rho_{s}\pm\sqrt{\sigma_{\rho_{s}\rho_{s}}}$}&\colhead{$r_{s}\pm\sqrt{\sigma_{r_{s}r_{s}}}$}&\colhead{$\sigma_{\rho_{s} r_{s}}$}&\colhead{$\Delta_{\mathrm{vir}}$\tablenotemark{a}}&\colhead{$r_{\mathrm{vir}}$}&\colhead{$(\chi^{2})	$\tablenotemark{b}}\\
   &\colhead{$\left[10^{15}h^{2}M_{\odot}\mathrm{Mpc}^{-3}\right]$}&\colhead{[Mpc/h]}&\colhead{$\left[10^{15}hM_{\odot}\mathrm{Mpc}^{-2}\right]$}&&\colhead{[Mpc/h]}&
   }
   \startdata
Abell~383&	$2.47\pm0.59$&	$0.33\pm0.04$&	-0.02&	111&	1.86&	2.0\\
Abell~209&	$1.14\pm0.29$&	$0.46\pm0.07$&	-0.02&	112&	1.95&	2.9\\
Abell~1423&	\nodata&		\nodata&		\nodata&	113&	\nodata&	\nodata\\	
Abell~2261&	$1.07\pm0.41$&	$0.51\pm0.11$&	-0.05&	114&	2.26&	3.7\\
RXJ2129+0005&	$2.16\pm0.67$&	$0.30\pm0.05$&	-0.04&	114&	1.65&	5.3\\
Abell 611&	$1.36\pm0.32$&	$0.41\pm0.06$&	-0.02&	118&	1.79&	4.1\\
MS2137-2353&	$1.14\pm0.20$&	$0.48\pm0.05$&	-0.01&	120&	1.89&	1.5\\
RXCJ2248-4431&	$1.24\pm0.34$&	$0.48\pm0.07$&	-0.02&	122&	1.92&	1.3\\
MACSJ1115+0129&	$0.61\pm0.17$&	$0.62\pm0.11$&	-0.02&	123&	1.78&	5.6\\
MACSJ1931-26&	$1.22\pm0.31$&	$0.41\pm0.07$&	-0.02&	123&	1.61&	4.2\\
RXJ1532.8+3021&	$1.16\pm0.52$&	$0.39\pm0.10$&	-0.05&	123&	1.47&	6.9\\
MACSJ1720+3536&	$2.44\pm0.84$&	$0.31\pm0.06$&	-0.05&	125&	1.61&	4.2\\
MACSJ0429-02&	$1.37\pm0.57$&	$0.41\pm0.08$&	-0.05&	126&	1.65&	1.9\\
MACSJ1206-08&	$2.60\pm0.94$&	$0.31\pm0.06$&	-0.05&	128&	1.63&	4.9\\
MACSJ0329-02&	$2.05\pm0.84$&	$0.33\pm0.08$&	-0.06&	129&	1.54&	6.3\\
RXJ1347-1145&	$2.10\pm0.90$&	$0.38\pm0.08$&	-0.07&	129&	1.80&	3.2\\
MACSJ1311-03&	$2.97\pm0.62$&	$0.24\pm0.03$&	-0.02&	131&	1.28&	4.0\\
MACSJ1423+24&	$3.70\pm1.83$&	$0.24\pm0.06$&	-0.11&	134&	1.34&	6.4\\
MACSJ0744+39&	$3.18\pm0.71$&	$0.28\pm0.04$&	-0.03&	141&	1.33&	3.2\\
CLJ1226+3332&	$3.72\pm0.83$&	$0.35\pm0.05$&	-0.04&	150&	1.57&	2.7\\
\enddata
\tablenotetext{a}{The virial overdensity at cluster redshift in units of the critical density.}
\tablenotetext{b}{The number of degrees of freedom is 10 in the case of CL~J1226 and 14 for all other clusters.} 
\end{deluxetable*}

\begin{deluxetable*}{lcccccccc}
 \tablecaption{NFW fits: mass-concentration parameters\label{NFW_Mc_Table}}
 \tablewidth{0pt}
\tablehead{
   \colhead{Name} &\colhead{$M_{2500c}$}&\colhead{$c_{2500c}$}&\colhead{$M_{500c}$}&\colhead{$c_{500c}$}&\colhead{$M_{200c}$}&\colhead{$c_{200c}$}&\colhead{$M_{\mathrm{vir}}$}&\colhead{$c_{\mathrm{vir}}$}\\
   &\colhead{$[10^{15}M_{\odot}/h]$}&&\colhead{$[10^{15}M_{\odot}/h]$}&&\colhead{$[10^{15}M_{\odot}/h]$}&&\colhead{$[10^{15}M_{\odot}/h]$}&
   }
   \startdata
Abell~383&	$0.26\pm0.05$&	$1.3\pm	0.3$&	$0.61\pm	0.07$&	$2.9\pm	0.7$&	$0.87\pm	0.07$&	$4.4\pm	1.0$&		$1.04\pm	0.07$&	$5.6\pm	1.3$\\
Abell~209&	$0.22\pm0.05$&	$0.9\pm	0.3$&	$0.63\pm	0.07$&	$2.1\pm	0.6$&	$0.95\pm	0.07$&	$3.3\pm	0.9$&		$1.17\pm	0.07$&	$4.3\pm	1.1$\\
Abell~1423&	\nodata&	\nodata&	\nodata&		  \nodata&		\nodata&	\nodata&   		\nodata&	    		\nodata     \\				
Abell~2261&	$0.34\pm0.12$&	$0.9\pm	0.4$&	$0.95\pm	0.16$&	$2.2\pm	0.9$&	$1.42\pm	0.17$&	$3.4\pm	1.4$&		$1.76\pm	0.18$&	$4.4\pm	1.8$\\
RXJ2129+0005&	$0.18\pm0.03$&	$1.2\pm	0.4$&	$0.43\pm	0.04$&	$2.8\pm	0.9$&	$0.61\pm	0.06$&	$4.3\pm	1.4$&		$0.73\pm	0.07$&	$5.6\pm	1.7$\\
Abell~611&	$0.21\pm0.04$&	$0.9\pm	0.3$&	$0.57\pm	0.04$&	$2.2\pm	0.6$&	$0.85\pm	0.05$&	$3.4\pm	0.9$&		$1.03\pm	0.07$&	$4.3\pm	1.1$\\
MS2137-2353&	$0.23\pm0.04$&	$0.8\pm	0.2$&	$0.68\pm	0.05$&	$2.0\pm	0.4$&	$1.04\pm	0.06$&	$3.1\pm	0.6$&		$1.26\pm	0.06$&	$4.0\pm	0.7$\\
RXCJ2248-4431&	$0.27\pm0.07$&	$0.8\pm	0.3$&	$0.76\pm	0.12$&	$2.0\pm	0.6$&	$1.16\pm	0.12$&	$3.2\pm	0.9$&		$1.40\pm	0.12$&	$4.0\pm	1.1$\\
MACSJ1115+0129&	$0.15\pm0.05$&	$0.5\pm	0.2$&	$0.54\pm	0.08$&	$1.4\pm	0.4$&	$0.90\pm	0.09$&	$2.3\pm	0.7$&		$1.13\pm	0.10$&	$2.9\pm	0.9$\\
MACSJ1931-26&	$0.16\pm0.03$&	$0.8\pm	0.2$&	$0.45\pm	0.04$&	$2.0\pm	0.6$&	$0.69\pm	0.05$&	$3.2\pm	0.9$&		$0.83\pm	0.06$&	$3.9\pm	1.1$\\
RXJ1532.8+3021&	$0.11\pm0.05$&	$0.8\pm	0.4$&	$0.34\pm	0.08$&	$1.9\pm	0.9$&	$0.53\pm	0.08$&	$3.0\pm	1.4$&		$0.64\pm	0.09$&	$3.8\pm	1.7$\\
MACSJ1720+3536&	$0.22\pm0.06$&	$1.2\pm	0.5$&	$0.53\pm	0.08$&	$2.8\pm	1.0$&	$0.75\pm	0.08$&	$4.3\pm	1.4$&		$0.88\pm	0.08$&	$5.2\pm	1.7$\\
MACSJ0429-02&	$0.19\pm0.11$&	$0.9\pm	0.4$&	$0.53\pm	0.13$&	$2.1\pm	0.9$&	$0.80\pm	0.14$&	$3.3\pm	1.3$&		$0.96\pm	0.14$&	$4.0\pm	1.6$\\
MACSJ1206-08&	$0.25\pm0.08$&	$1.2\pm	0.5$&	$0.60\pm	0.11$&	$2.8\pm	1.0$&	$0.86\pm	0.11$&	$4.3\pm	1.5$&		$1.00\pm	0.11$&	$5.2\pm	1.7$\\
MACSJ0329-02&	$0.20\pm0.06$&	$1.1\pm	0.4$&	$0.50\pm	0.09$&	$2.5\pm	1.1$&	$0.73\pm	0.10$&	$3.8\pm 1.6$&		$0.86\pm	0.11$&	$4.7\pm	1.9$\\
RXJ1347-1145&	$0.31\pm0.13$&	$1.1\pm	0.5$&	$0.79\pm	0.19$&	$2.5\pm	1.1$&	$1.16\pm	0.19$&	$3.9\pm	1.5$&		$1.35\pm	0.19$&	$4.7\pm	1.8$\\
MACSJ1311-03&	$0.14\pm0.02$&	$1.3\pm	0.3$&	$0.32\pm	0.19$&	$2.9\pm	0.6$&	$0.46\pm	0.03$&	$4.4\pm	1.0$&		$0.53\pm	0.04$&	$5.3\pm	1.1$\\
MACSJ1423+24&	$0.18\pm0.08$&	$1.4\pm	0.8$&	$0.41\pm	0.06$&	$3.1\pm	0.8$&	$0.57\pm	0.10$&	$4.7\pm	1.2$&		$0.65\pm	0.11$&	$5.7\pm	2.8$\\
MACSJ0744+39&	$0.20\pm0.03$&	$1.2\pm	0.3$&	$0.49\pm	0.04$&	$2.7\pm	0.6$&	$0.70\pm	0.04$&	$4.1\pm	1.0$&		$0.79\pm	0.04$&	$4.8\pm	1.1$\\
CLJ1226+3332&	$0.43\pm0.07$&	$1.1\pm	0.3$&	$1.08\pm	0.09$&	$2.6\pm	0.6$&	$1.56\pm	0.10$&	$4.0\pm	0.9$&		$1.72\pm	0.11$&	$4.5\pm	1.1$\\
\enddata

\end{deluxetable*}

\section{General Concentration-Mass Analysis}\label{GMc}
We now derive a concentration-mass relation from our 19 X-ray selected CLASH clusters and compare the observed values to the theoretical expectations from the literature. 
In the following, we will quote mass and concentration values which refer to a halo radius of $r_{200\mathrm{c}}$.

\subsection{The Concentration-Mass Relation from CLASH}\label{CLASH_c_m}

\begin{figure}
 \begin{center}
  \includegraphics[width=.48\textwidth]{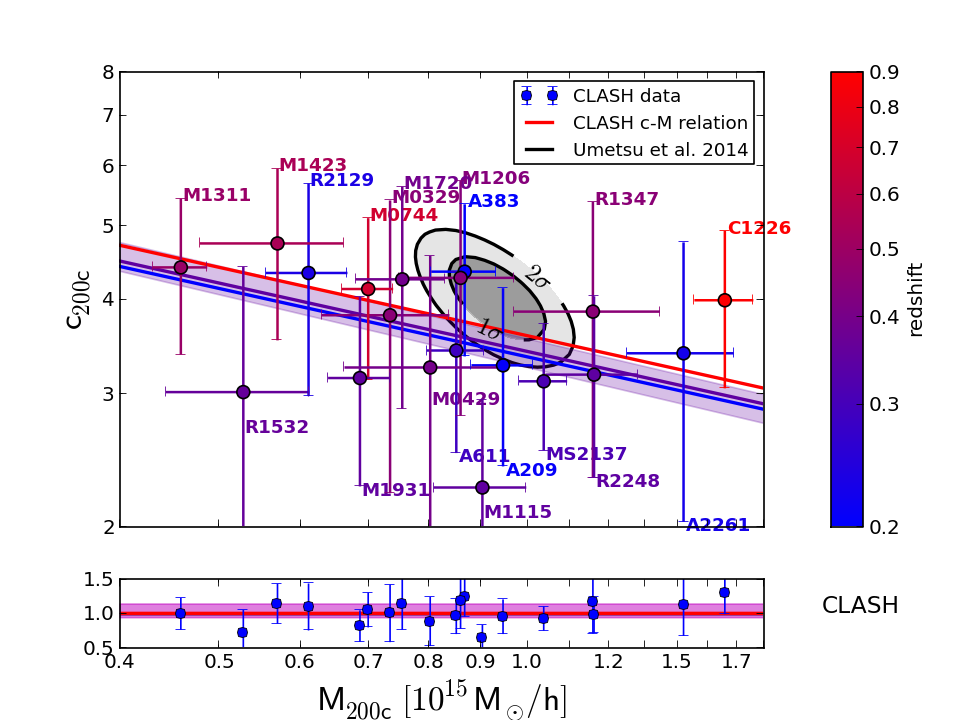}\\
 \end{center}
  \caption{Concentrations and masses from CLASH. The labeled data points in the \textit{top panel} show each CLASH cluster in the $M_{200c}-c_{200c}$ plane.
  The solid lines show the best-fit c-M relation to the CLASH data for $z=0.2$ (blue), $z=0.35$ (purple) and $z=0.9$ (red). The color of data points
   and lines encodes the redshift of the CLASH clusters or the c-M relation. The shaded band around the central ($z=0.35$) relation
   shows the $1\sigma$ error on the normalization of the CLASH c-M relation.
   Overplotted with the gray contours is the 
   concentration-mass analysis of \citet{Umetsu2014} for a redshift of $z\simeq 0.35$. The contour lines
   encircle the 68\% and 95\% confidence levels, respectively.
   The \textit{bottom panel} shows the ratio
   between the observed concentration value and the value predicted by the CLASH-derived c-M relation for each CLASH cluster.
   The red line shows the median of this ratio for all clusters and the pink area defines
   the interval between its first and third quartile. \label{c_m_CLASH}.}
\end{figure}

In Fig.~\ref{c_m_CLASH} we visualize the CLASH data points from Table~\ref{NFW_Mc_Table} in the c-M plane. A general statistical summary of the data is shown in Table~\ref{c_m_men_CLASH_tab}. In order to derive a c-M relation, we choose a parametrization following \citet{Duffy2008}, but with pivot mass and redshift matched to our sample,
\begin{eqnarray}
c_{200c}(M_{200c},z)&=&A\times\left(\frac{1.37}{1+z}\right)^B \nonumber \\ 
&\times&\left(\frac{M_{200c}}{8\times10^{14}M_{\odot}/h}\right)^C \; .
\label{eqn:cmrel}
\end{eqnarray}
Here, $A$ is the concentration of a halo at the pivot mass and redshift, $B$ the redshift dependence of the concentration and $C$ the dependence on halo mass.

Our data used in the fit contain errors in both mass and concentration, and we expect an intrinsic scatter about the mean relation. Despite this, unbiased estimates of the parameters of the relation can be determined using a likelihood method \citep[e.g.][]{Kelly2007}. In analogy to \citet{Hoekstra2012} and \citet{Gruen2013}, we write the likelihood with an additional term that includes the intrinsic scatter as
\begin{eqnarray}
-2\ln\mathcal{L}&=& \sum_i\ln(s_i^2)+ \left(\frac{\ln(c_i)-\ln(c(M_i,z_i))}{s_i} \right)^2 \nonumber \\
s_i^2&=&C^2\sigma^2_{\ln M, i}+\sigma^2_{\ln c, i}+\sigma^2_{\ln c,\mathrm{int}} \; ,
\end{eqnarray}
where we use the single-parameter ln-normal measurement uncertainties of mass and concentration $\sigma_{\ln M, i}$ and $\sigma_{\ln c, i}$, an intrinsic ln-normal scatter of concentration $\sigma_{\ln c,\mathrm{int}}$ and Eqn.~\ref{eqn:cmrel} as $c(M_i,z_i)$, with a sum over all clusters $i$. The likelihood is a function of both the parameters $A,B,C$ and $\sigma_{\ln c,\mathrm{int}}$. For our measurements, it is maximized by $A=3.66\pm0.16$, $B=-0.14\pm0.52$ and $C=-0.32\pm0.18$, where the errors are close to uncorrelated. The intrinsic scatter is consistent with zero at a $1\sigma$ upper limit of $\sigma_{\ln c,\mathrm{int}}=0.07$. 

The results can be summarized as follows:
\begin{itemize}
\item The concentration at the mean mass and redshift of the CLASH sample is constrained at the 5\% level.
\item We detect a trend towards lower concentration at higher mass with moderate significance, in agreement with theoretical expectations \citep{Duffy2008,Bhattacharya2013}.
\item Due to the limited dynamic range, our data allow no conclusion on the dependence of concentration on redshift. The theoretical expectation here is to find a negative dependence on redshift from the combined effect of density at the formation time and mass growth \citep[e.g][]{Navarro1997,Duffy2008}.
\end{itemize}

To confirm our result with another concentration-mass analysis, which is of course not fully independent but different in its
methodology, we overplot in Fig.~\ref{c_m_CLASH} the c-M contours at the 68\% and 95\% confidence levels from \citet{Umetsu2014}.
These contours derive from the stacked weak-shear analysis of 16 CLASH X-ray selected clusters. Although the stacked result, which 
refers to a redshift of $z\simeq 0.35$, lies slightly above the value from our relation, it is in excellent agreement with our
results given the uncertainties in both analyses.

\subsection{Comparison with Results from the Literature}
We choose the relations 
of \citet{Duffy2008} (hereafter D08) and \citet{Bhattacharya2013} (hereafter B13) for our comparison to the CLASH data.

D08 used a set of three N-body simulation runs with a co-moving box size of 25, 100 and 400 Mpc/h, respectively. All runs adopted a WMAP5 cosmology \citep{Komatsu2009} and provided a total mass-range of $10^{11}$--$10^{15}M_{\odot}/h$.
In addition, D08 also defined a relaxed sub-sample, with the criterion that 
the separation between the most bound halo particle and the center of mass of the halo is smaller than $0.07r_{\mathrm{vir}}$, which 
was formerly identified as one efficient way of selecting relaxed halos \citep{Neto2007}.

B13  worked with a set of four cosmological boxes ranging in co-moving box size from 128--2000 Mpc/h. 
Also B13 splits their sample into a full and a relaxed subset, where the relaxed one
is defined by the same criterion as in D08. Apart from the larger cosmological boxes, the main difference between D08 and B13 is
the cosmological background model, which resembled a WMAP7 \citep{Komatsu2011} cosmology in the case of B13 and the larger box size.
\subsubsection{c-M Relation of the Full Samples}
\begin{figure}
 \begin{center}
  \includegraphics[width=.48\textwidth]{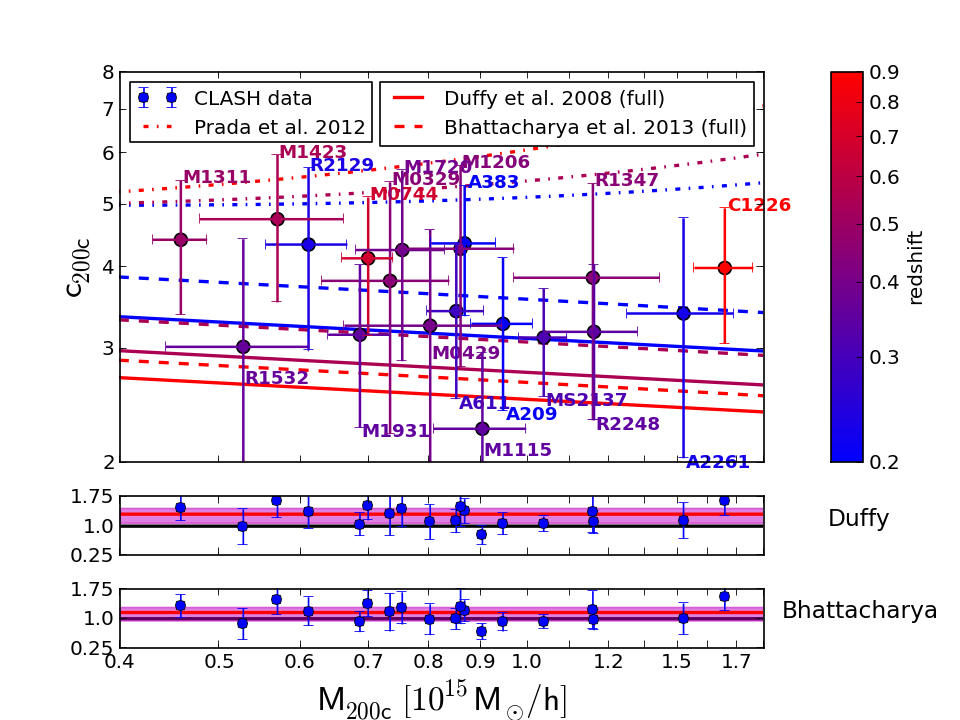}\\
 \end{center}
   \caption{A comparison between CLASH clusters and c-M relations from the literature. This figure is identical in its structure
   to Fig.~\ref{c_m_CLASH}. The lines indicate the c-M relations for the full samples of D08, B13 and P12. The \textit{bottom
   panels} show the ratio of the observed and the simulation-based concentration, together
   with the sample median of this ratio and its quartiles. \label{c_m_full}}
\end{figure}

First, we compare the CLASH data set to the full sample c-M relations of D08 and B13. 
As one can see from visual inspection of Fig.~\ref{c_m_full} already, there is good agreement between the 
CLASH data and the theoretical c-M relations derived from the simulations.

To statistically quantify the agreement we calculate the ratio $c_{\mathrm{obs}}/c_{\mathrm{sim}}$ as a function of cluster redshift.
This ratio for each data point is shown in the bottom panel of Fig.~\ref{c_m_full}. Next, we calculate the mean, standard deviation, 
first, second (median) and third quartiles of all these ratios and report them in Table~\ref{c_m_full_tab}. The median is also shown as horizontal
pink line in the bottom panel of Fig.~\ref{c_m_full} with the error range defined by the first and third quartiles. 
As a last test we perform a Pearson's $\chi^{2}$ test, with the null hypothesis that the theoretical c-M relation is a good fit
to our data and report the p-value in Table~\ref{c_m_full_tab}. All the analysis components, described in this paragraph shall
serve as the prototype for all following comparisons between our data and c-M relations from simulations.
To quantify how well we can expect the data and c-M relation to agree, we show in the very top of Table~\ref{c_m_full_tab}
the comparison to the c-M relation which we derived in Sec.~\ref{CLASH_c_m} from the CLASH data itself. 

Finally, Fig.~\ref{c_m_full}
also shows the c-M relation of \citet{Prada2012} since it is widely used in the literature. One can easily see that there is 
a discrepancy between this relation and the CLASH results, especially when the good agreement with the D08 and B13 relations 
is considered. However, we refer to \citet{Meneghetti2013} which argues that a
direct comparison in the c-M plane is not a meaningful comparison in the case of the \citet{Prada2012} relation.

\begin{deluxetable}{ccccccc}
 \tablecaption{Goodness-of-fit: CLASH compared to literature samples\label{c_m_full_tab}}
 \tablewidth{0pt}
 \tablehead{
   \colhead{Reference} &\colhead{$\left<c_{\mathrm{obs}}/c_{\mathrm{sim}}\right>$\tablenotemark{a}}&\colhead{$Q_{2}$\tablenotemark{b}}&\colhead{$Q_{1}$\tablenotemark{c}}&\colhead{$Q_{3}$\tablenotemark{d}}&\colhead{$\chi^{2}$}&\colhead{p-value}\\
}
\startdata
   CLASH c-M&$1.02\pm0.17$&1.01&0.94&1.14&7.6&0.94\\
   D08 (full)&$1.26\pm0.24$&1.31&1.07&1.45&15.3&0.43\\
   B13 (full)&$1.12\pm0.23$&1.16&0.94&1.29&11.4&0.72\\
   D08 (relaxed)&$1.11\pm0.21$&1.15&0.95&1.27&10.1&0.81\\
   B13 (relaxed)&$1.08\pm0.23$&1.12&0.91&1.24&11.3&0.73\\
\enddata
  \tablenotetext{a}{The mean of $c_{\mathrm{obs}}/c_{\mathrm{sim}}$ for the full cluster sample.}
  \tablenotetext{b}{The second quartile or median.}
  \tablenotetext{c}{The first quartile (25\%).}
  \tablenotetext{d}{The third quartile (75\%).}
\end{deluxetable}

\subsubsection{c-M Relation for the Relaxed Samples}
\begin{figure}
 \begin{center}
  \includegraphics[width=.48\textwidth]{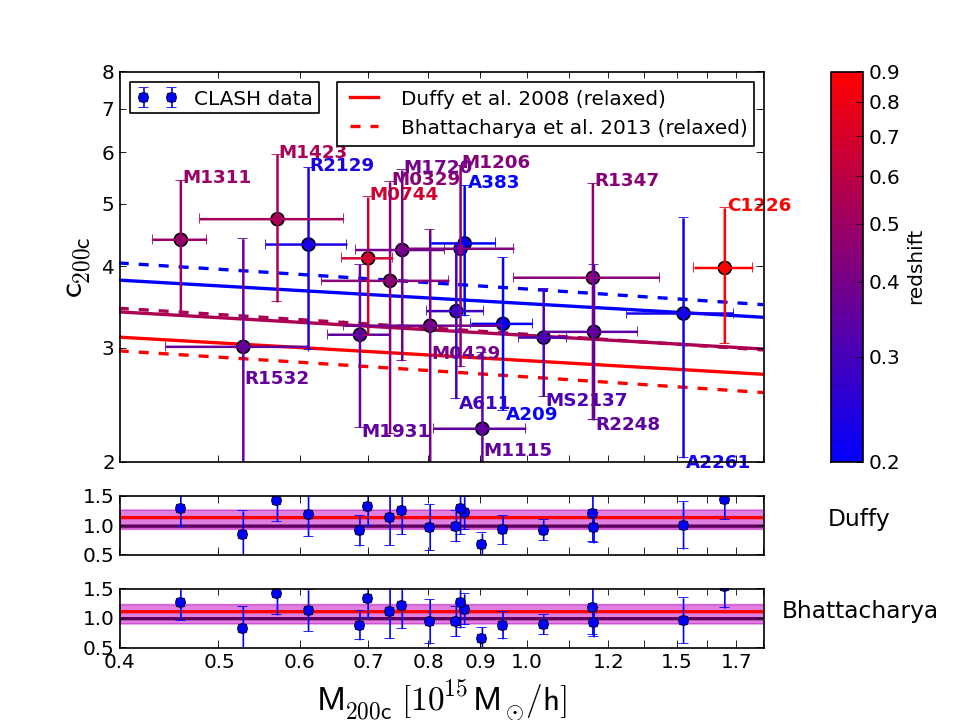}\\
 \end{center}
  \caption{This figure is identical to Fig.~\ref{c_m_full} but shows the c-M relations derived from the relaxed samples
  of D08 and B13.\label{c_m_relaxed}}
\end{figure}

Since the CLASH clusters were selected to represent a more relaxed sample of clusters than former studies, 
we expect a much higher level of agreement when comparing to the relaxed subsets of 
the simulations. The visual comparison is shown in Fig.~\ref{c_m_relaxed}, together with the statistical assessment in
Table~\ref{c_m_full_tab}. The expectation of a better agreement is indeed satisfied, especially in the case of D08 where a 31\% 
difference between simulation and observation is reduced to a 15\% difference. 
Note that the change from the full to the relaxed sample c-M relation
in the work of B13 is only marginal (from 16\% difference to 12\%), although the same relaxation criterion was applied as in D08. This might either
be caused by the different cosmology used in the two simulations or might relate to the much bigger set of clusters in B13 and 
the increased statistical power of the sample.

\section{Concentration-Mass Analysis with a Tailored Set of Simulations}\label{SMc}
In the preceding analysis we ignored the fact that we derive our NFW fits from a lensing reconstruction which sees the clusters in projection and we have not properly accounted 
for the CLASH selection function. We aim at eliminate these shortcomings by using our own set of simulations, where we have full control over the selection of our
halo sample and the way in which masses and concentrations are derived from the simulations.

In our companion paper \citet{Meneghetti2014} (hereafter M14) we use a set of 1419 cluster-sized halos from the MUSIC-2 sample \citep{Sembolini2013}. These halos were found in the 1 Gpc MultiDark cosmological simulation \citep{Klypin2011,Riebe2013}
which was run with a best-fit WMAP7+BAO+SNI cosmology ($\Omega_{M}=0.27$, $\Omega_{\Lambda}=0.73$ h=0.7). Starting from the large cosmological box with coarse particle mass resolution, the zoom-technique \citep{Klypin2001} was applied to run
re-simulations of the halos of interest with added non-radiative gas physics. 
This comprehensive 
set of clusters spans a mass-range between $2\times10^{14}M_{\odot}/h$ -- $2\times10^{15}M_{\odot}/h$ at $z=0$ and is available at four different redshifts (0.25,~0.33,~0.43,~0.67).
More details on this set of numerically simulated clusters are given in M14 and Vega et al.~(2014 in prep).

\subsection{Analysis in 3D}\label{M14_3D}
We measure masses and
concentrations of the halos in our simulated sample in a standard way by counting particles in radial bins around the halo center 
and by assigning a mean density to each bin. The innermost radial bin in this scheme is defined by the spatial resolution
of the underlying zoom simulations and the outermost radial bin refers to $r_{200\mathrm{c}}$ of the halo. We fit a NFW profile to the
decadic logarithm of the density as described in more detail in \citet{Ludlow2013} and M14. To the measured
masses and concentrations of each halo and at all available redshifts we fit a c-M relation following the parametrization of \citet{Duffy2008}, adapted to the mass and redshift range of the simulations.

\begin{figure}
 \begin{center}
  \includegraphics[width=.48\textwidth]{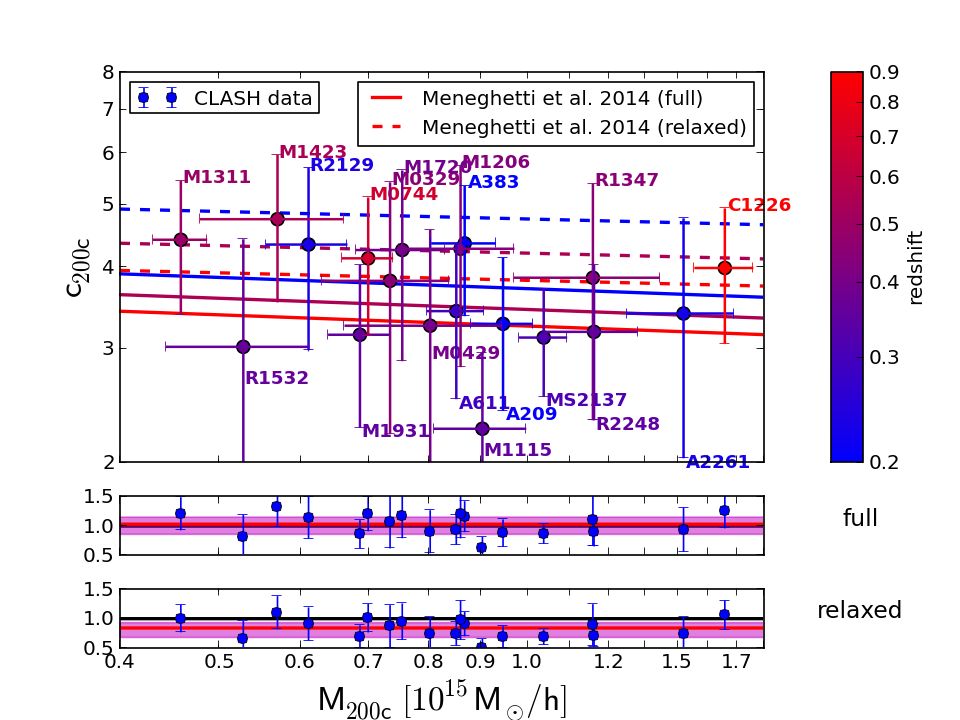}\\
 \end{center}
  \caption{A comparison between CLASH and a tailored set of c-M relations from numerical simulations. This figure is identical in its
  structure to Fig.~\ref{c_m_full} and shows the comparison between the CLASH data and the analysis of the simulations by
  M14 in 3D.\label{c_m_men_3D}}
\end{figure}

To define a limiting case we construct a strictly relaxed subset\footnote[1]{This is defined as the ``super-relaxed'' sample in M14} of our simulated sample,
by applying all three relaxation criteria of \citet{Neto2007}. In addition to the already mentioned ratio of center of mass and
virial radius, this includes also a constraint on the halo's substructure mass fraction $f_{\textrm{sub}}<0.1$ and the 
restriction that the virial ratio must obey $2T/|U|<1.35$. For complete definitions of $f_{\textrm{sub}}$, $T$ and $U$ see \citet{Neto2007}.
This selection reduces the number of halos in this strictly relaxed subset to 15\% of the original full sample. Please note that this
relaxation criterion is indeed more restrictive than the one used by D08 and B13 which only obeyed the center of mass constraint.
The c-M relations for both the full and the relaxed sample are shown 
in Fig.~\ref{c_m_men_3D}.

We summarize the quantitative comparison to our observed CLASH results in Table~\ref{c_m_men_tab}. We find excellent agreement between our observed data and the full sample of M14, very similar
to the findings of B13. It is indeed reassuring that our baseline c-M relation derived from the full set of simulated clusters
and analyzed with standard profile-fitting techniques gives a very similar result to B13 since most of our sample clusters 
were selected from the same parent cosmological box. The picture changes however,
when we turn our attention to the strictly relaxed sample of M14, as can be seen in Fig.~\ref{c_m_men_3D} and Table~\ref{c_m_men_tab}. 
On average, the concentrations of the CLASH sample are underestimated by about 15\% and the 
associated p-value drops from 0.85 in the full to about 0.01 in the relaxed sample. 
This is in some tension with the results seen for D08 and B13, 
but we remind the reader that the selection criteria we adopt differ from those in D08 and B13. Specifically, 
we adopt all three criteria as used by
\citet{Neto2007} to create the limiting
case of a strictly relaxed sample, while D08 and B13 used a less strict definition of relaxation based on only one of these criteria. 

\subsection{Analysis in 2D}\label{M14_2D}
One aspect of our analysis may introduce substantial biases, namely
that we assume spherical symmetry while fitting a 3D radial profile to our projected data coming from a lensing reconstruction.
Several solutions to work around this issue have been proposed, e.g. by using X-ray and SZ data to gain information on the 3D shape
of the density profile \citep[see e.g.][]{Mahdavi2007, Corless2009, Morandi2010, Morandi2012, Sereno2013}. In this work we choose a different
approach by also analyzing our simulated data in projection and by making the same assumption of spherical symmetry when deriving 
the density profiles of the simulated halos. 

We perform the projection for each of our halos in the full sample by projecting all simulation particles in a box with 6 Mpc/h sides
around the halo center. We chose 100 randomly selected lines-of-sight to obtain many realizations of the same
halo, thereby increasing the statistical power of our sample. From the projected particle density we derive convergence maps, bin them
azimuthally around the halo center and fit a NFW profile to the binned data under the assumption of spherical symmetry. For more details
we refer the interested reader to M14 and Vega et al.~(2014 in prep.). Also for this 2D case, we define a strictly relaxed sample as limiting case 
following the criteria outlined in 
Sec.~\ref{M14_3D}.

\begin{figure*}
 \begin{center}
  \includegraphics[width=.98\textwidth]{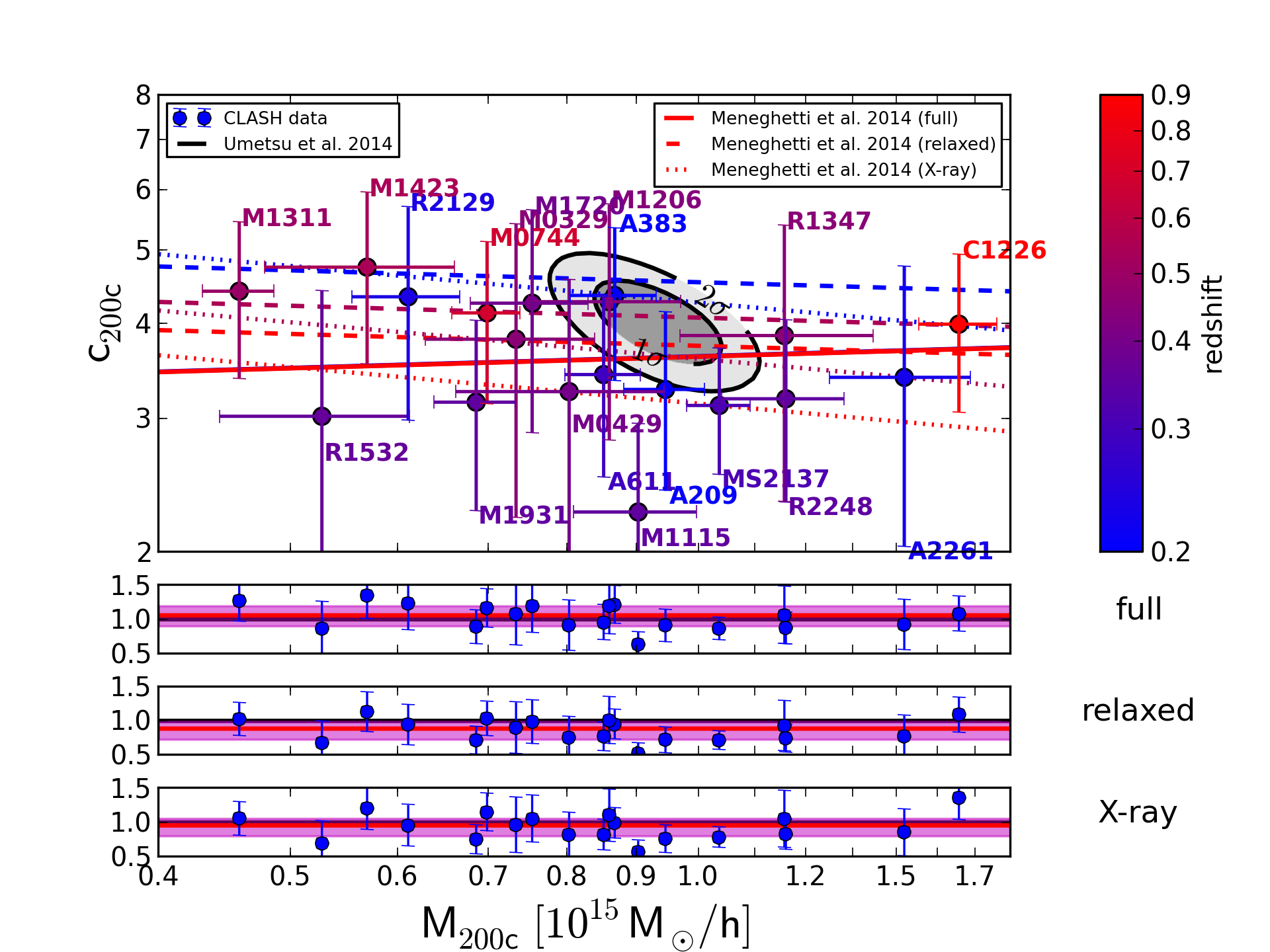}\\
 \end{center}
   \caption{A c-M comparison in 2D. This figure is identical to Fig.~\ref{c_m_men_3D}, but shows the comparison between
   different c-M relations, based on different halo subsets from M14 in 2D. In addition, we overlay again the c-M likelihood
   contours from \citet{Umetsu2014}.\label{c_m_men_2D}}
\end{figure*}

The results of the comparison to these 2D c-M relations with the CLASH data can be seen in Fig.~\ref{c_m_men_2D}. By applying the
same statistical tests we find an excellent agreement with the full 2D sample of M14. 
On average, the observed concentrations are only 6\% higher than in the simulated sample which is now free of the projection bias.
However, when restricting ourselves to the strictly relaxed clusters the 2D c-M relation is in tension with observations.
The p-value drops from 0.87 to 0.01 and the difference in the median concentration ratio increases to 12\%. Furthermore, the trend in the difference has changed. The expected values from the simulations are now higher than the observations.
The situation improves significantly to only 7\%  overestimation in the concentration ratio and a p-value of 0.26 once we pick only those simulated clusters which are able to produce strong-lensing
features by demanding that the cluster produces a critical line (comp.~Eq.~\ref{Jacobiandeterminant}). This selection is appropriate 
since all but one CLASH cluster allowed us to identify strong lensing features. However, the observational data is clearly
in tension with a simulated cluster sample selected after the three relaxation criteria of \citet{Neto2007} and which
is analyzed in 2D. This highlights the importance of halo selection and the necessity to properly account for the CLASH selection function. 

\subsubsection{X-ray Selection of CLASH Clusters}
As is pointed out in \citet{Postman2012}, the CLASH X-ray selected sample was designed to have a mostly regular X-ray morphology. 
Therefore, we perform yet another selection from our M14 cluster sample, mimicking the CLASH 
X-ray selection. As pointed out in Sec.~6 of M14, the selection based on X-ray regularity is related to but not identical 
to a selection based on halo relaxation.
The X-ray selection is possible with the help of the X-MAS simulator \citep{Rasia2008,Meneghetti2010a,Rasia2012} which produces
simulated X-ray observations from a numerically simulated halo. We configure X-MAS to reproduce the 
X-ray observations \citep{Maughan2008,Allen2008,Ebeling2007,Cavagnolo2008,Mantz2010} according to which the CLASH clusters  were selected.
Using this set of simulated X-ray images
we apply the very same selection criteria which were used to select the CLASH X-ray selected clusters. 
For a more detailed description of these criteria and the selection process see M14.

This CLASH-like, X-ray selected sample in 2D is the one simulated sample which comes closest to the real CLASH clusters, both 
with respect to the selection criteria and the analysis method. The comparison between the c-M relation of this sample
and the observed CLASH clusters shows indeed significant improvement over the limiting case of the fully relaxed sample in the last section.
The qualitative agreement between the data points and the X-ray selected c-M relation in Fig.~\ref{c_m_men_2D} is quite obvious.
The median concentration ratio shows that the observed CLASH concentrations are only 4\% lower than the ones
from the X-ray selected simulation sample and the p-value 0.38 indicates no strong tension between the two samples (compare Table~\ref{c_m_men_tab}).
Finally, we calculate the $\Delta\chi^{2}$ value for the fits of the CLASH c-M relation from Sec.~\ref{CLASH_c_m} and the X-ray
selected c-M relation and find that the two relations agree at the 90\% confidence level. 
\begin{deluxetable}{lcccccc}
 \tablecaption{Goodness-of-fit: Meneghetti et al.~2014\label{c_m_men_tab}}
\tablewidth{0pt}
\tablehead{
   \colhead{Sample} &\colhead{$\left<c_{\mathrm{obs}}/c_{\mathrm{sim}}\right>$}&\colhead{$Q_{2}$}&\colhead{$Q_{1}$}&\colhead{$Q_{3}$}&\colhead{$\chi^{2}$}&\colhead{p-value}
}
\startdata
   3D full 		&$1.00\pm0.18$&1.03&0.86&1.15&9.5&0.85\\
   3D relaxed		&$0.80\pm0.16$&0.84&0.68&0.93&29.4&0.01\\
   2D full		&$1.03\pm0.19$&1.06&0.89&1.09&9.2&0.87\\
   2D relaxed		&$0.86\pm0.16$&0.88&0.73&0.98&32.1&0.01\\
   2D SL		&$0.91\pm0.19$&0.93&0.78&1.03&18.0&0.26\\
   2D X-ray		&$0.94\pm0.20$&0.96&0.80&1.06&16.1&0.38\\
\enddata
\tablecomments{The column explanations are identical to Table~\ref{c_m_full_tab}.}
\end{deluxetable}

\subsection{Individual CLASH Clusters in Our Simulated Sample}
\begin{figure}
 \begin{center}
  \includegraphics[width=.48\textwidth]{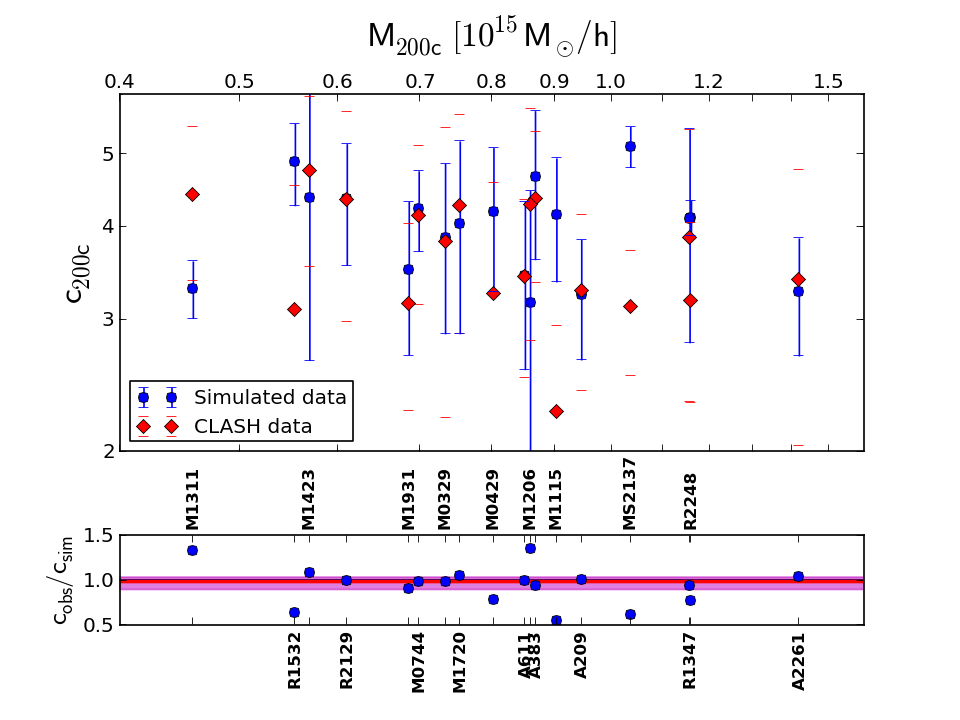}
 \end{center}
  \caption{The distribution of observed and simulated concentrations for 18 X-ray selected CLASH clusters. The blue points show 
  the expected concentration for each CLASH cluster as it is derived from all halo projections which fulfill our
   CLASH X-ray selection criteria for that specific halo. The red points show the observational equivalent. In the 
   \textit{bottom panel} we show the ratio between the two concentration values, together with the median of the ratio sample in red.
   The pink error band is defined by the first and third quartile of this sample. \label{c_m_men_CLASH}}
\end{figure}

As the final analysis in this work we now select close matches to individual CLASH clusters out of our 2D set of simulated halos. 
We do this in order to gather additional confirmation that our specific way of selecting CLASH clusters from a numerical simulation
is sufficiently accurate to characterize the CLASH selection function.
We find simulated counterparts to individual CLASH clusters
by means of a regularity metric defined in Sec.~4 of M14. 
After all matching projections have been found for a single CLASH cluster, we
calculated the expected concentration by a weighted average over the concentrations of these projections (see Sec.~7 of M14 for details). 
In the course of this analysis we had to drop CL~J1226  because no match was found in our simulated set. The system is very massive and sits at high redshift which would require a larger simulation to find an 
equivalent\footnote[1]{An even more massive system at similar redshift has been observed \citep[e.g.][]{Jee2013,Menanteau2012}.}. 
We show the findings of the remaining 18 systems in Fig.~\ref{c_m_men_CLASH}, where we compare the expected concentration for
each individual simulated CLASH-like cluster with the findings from observations. All but two points overlap within
the 1$\sigma$ error bars and the ratio between observed and simulated concentrations for all CLASH clusters is close to a perfect 
match with the median of $c_{\mathrm{obs}}/c_{\mathrm{sim}}$ $Q_{2}=0.99^{+0.05}_{-0.09}$ where the error margins are defined
by the first and third quartiles of the sample. The fact that the selection of individual CLASH clusters shows
good agreement between predicted and observed concentrations gives us some confidence that we are	 indeed able 
to characterize the CLASH selection function by means of X-ray morphology. 

We provide a general statistical summary of 
the distribution of simulated concentrations in Table~\ref{c_m_men_CLASH_tab} and we conclude
our comparison to the simulations of M14 with a two-sample statistical analysis. 
We perform a Kolmogorov-Smirnov test and find a p-value of 0.75, 
again showing no indication for tension in the null hypothesis that the observed and 
simulated data have the same parent distribution of c-M values.

\begin{deluxetable}{lccccccc}
 \tablecaption{General properties of concentration samples\label{c_m_men_CLASH_tab}}
\tablewidth{0pt}
\tablehead{
   \colhead{Sample} &\colhead{mean}&\colhead{SD}&\colhead{$Q_{2}$}&\colhead{$Q_{1}$}&\colhead{$Q_{3}$}&\colhead{min}&\colhead{max}
 }
 \startdata
   Observed data&3.65&0.65&3.43&3.18&4.26&2.26&4.75\\
   Simulated data&3.87&0.61&3.76&3.62&3.93&3.07&5.68\\
\enddata
\tablecomments{The column explanations are identical to Table~\ref{c_m_full_tab}.}
\end{deluxetable}

\section{Conclusions}
\label{Concl}
The HST multi-cycle treasury program CLASH was in part designed to shed light on the dark matter density profile
of galaxy clusters by combining the enormous resolving power of the HST with wide-field Subaru imaging. 
The CLASH X-ray selected sample of galaxy clusters was specifically selected to have a mostly undisturbed X-ray morphology, suggesting that this sub-sample
represents an undisturbed and unbiased set of clusters in terms of their density profile. 
This choice was made since former studies
of lensing clusters with exquisite data quality 
were inconsistent with the predictions of $\Lambda$CDM, and selection effects were thought to be a possible 
cause of this disagreement.

In this work we applied advanced lensing reconstruction techniques to this CLASH data set.  
Our reconstructions combines weak and strong lensing to fully exploit
the lensing data  provided by the CLASH program.
With the help of adaptively refined grids, we achieve a non-parametric reconstruction of the lensing
potential over a wide range of scales, from the inner-most strong-lensing core of the system at scales 
$\sim 10$kpc out to the virial radius at $\sim 2$ Mpc. This is the first time that such a multi-scale
reconstruction using weak and strong lensing has been performed for such a large sample of clusters. 
Fits to the surface-mass density profiles provide
masses and concentrations for 19 massive galaxy clusters.

In order to have full control over the selection function of halos and in order to avoid possible biases introduced 
by the tri-axial structure of high-mass halos, we also derive c-M relations from a new, unique set of 
simulated  halos. These simulations allow us to make specific choices in our selection and analysis, providing a much closer
match to real observations. While simulations
are usually analyzed in 3D  we perform a purely 2D analysis in projection,
as this is the only option for the observed lensing data. We apply different selection functions to the simulations, including
a selection based on the X-ray morphology of realistic X-ray images of our hydro-simulations. This sample obeys the selection
criteria of CLASH. This is of great importance since the selection of a cluster from a numerical simulation based on X-ray regularity, like in
the case of CLASH, relates to but is not identical to a selection based on relaxation parameters only. 
The details of this selection function are studied in much more detail in another CLASH paper by \citet{Meneghetti2014}. 
For the X-ray selected 2D sample we find excellent agreement between simulations and observations.  Observed concentration are on average
only 4\% lower than in simulations and we find no statistical indication
for tension between the simulated and observed data set.
This detailed comparison between observations and simulations in 2D, with full consideration of the underlying selection
function is unique and gives us great confidence in the results, which are a confirmation of the $\Lambda$CDM paradigm, at 
least in the context of a c-M relation of cluster-sized halos. 

From fitting a c-M relation to the CLASH data directly we find our concentrations distributed around a central value
of $c_{200\textrm{c}}\simeq3.7$ with a mild negative trend in mass at the $1\sigma$-level.
This c-M relation derived from the CLASH data directly agrees with the c-M relation of simulated X-ray selected halos analyzed in projection
at the 90\% confidence level.
Our comprehensive likelihood analysis shows that we are insensitive to any possible redshift dependence of the 
c-M relation. A larger leverage in redshift would be needed to probe this trend which is suggested by numerical simulations.

We want to highlight the complementary work on  CLASH weak lensing and magnification measurements by \citet{Umetsu2014}
and the full characterization of the CLASH simulations by \citet{Meneghetti2014}. However,
due to the exquisite quality of the lensing data used for this analysis, further and more advanced studies will be possible.
Ongoing analyses include additional functional forms describing the dark matter distribution, like the
generalized NFW or Einasto profiles. Particularly the analysis of inner slopes of the CLASH clusters and the intrinsic scatter of 
c-M relations derived from these profiles
will give interesting insights into the physics of dark matter and the role of baryons on cluster scales. Ultimately, one would like
to go away from 1D, radial density profiles and describe the full morphology and shape of the dark matter distributions in observations 
and simulations. Such techniques might indeed prove more powerful in e.g.~distinguishing different particle models of dark matter. 
The CLASH clusters are clearly the ideal data set to perform such analyses.

\acknowledgments
The research was in part carried out at the Jet Propulsion Laboratory,
California Institute of Technology, under a contract with the National Aeronautics and Space Administration.
M.M.~thanks ORAU and NASA for supporting his research at JPL and acknowledges 
support from the contract ASI/INAF I/023/12/0, INFN/PD51, and the PRIN MIUR 2010Ð2011 
''The dark Universe and the cosmic evolution of baryons: from current surveys to Euclid''.
K.U.~acknowledges support from the National Science Council of Taiwan (grant NSC100-2112-M-001-008-MY3) and from the Academia Sinica Career Development Award.
Support for A.Z.~is provided by NASA through Hubble Fellowship grant \#HST-HF-51334.01-A awarded by STScI.
D.G., S.S.~and P.R.~were supported by SFB Transregio 33 'The Dark Universe' by the Deutsche Forschungsgemeinschaft (DFG) 
and the DFG cluster of excellence 'Origin and Structure of the Universe'.
This work was supported in part by contract research ``Internationale
Spitzenforschung II/2-6'' of the Baden W\"{u}rttemberg Stiftung.
The Dark Cosmology Centre is funded by the DNRF.
J.S.~was supported by NSF/AST1313447, NASA/NNX11AB07G, and the Norris
Foundation CCAT Postdoctoral Fellowship. E.R. acknowledges support from the
National Science Foundation AST-1210973, SAO TM3-14008X (issued under NASA Contract No. NAS8-03060)


\newpage

\appendix
\begin{figure*}[!h]
 \begin{center}
 \includegraphics[width=.85\textwidth]{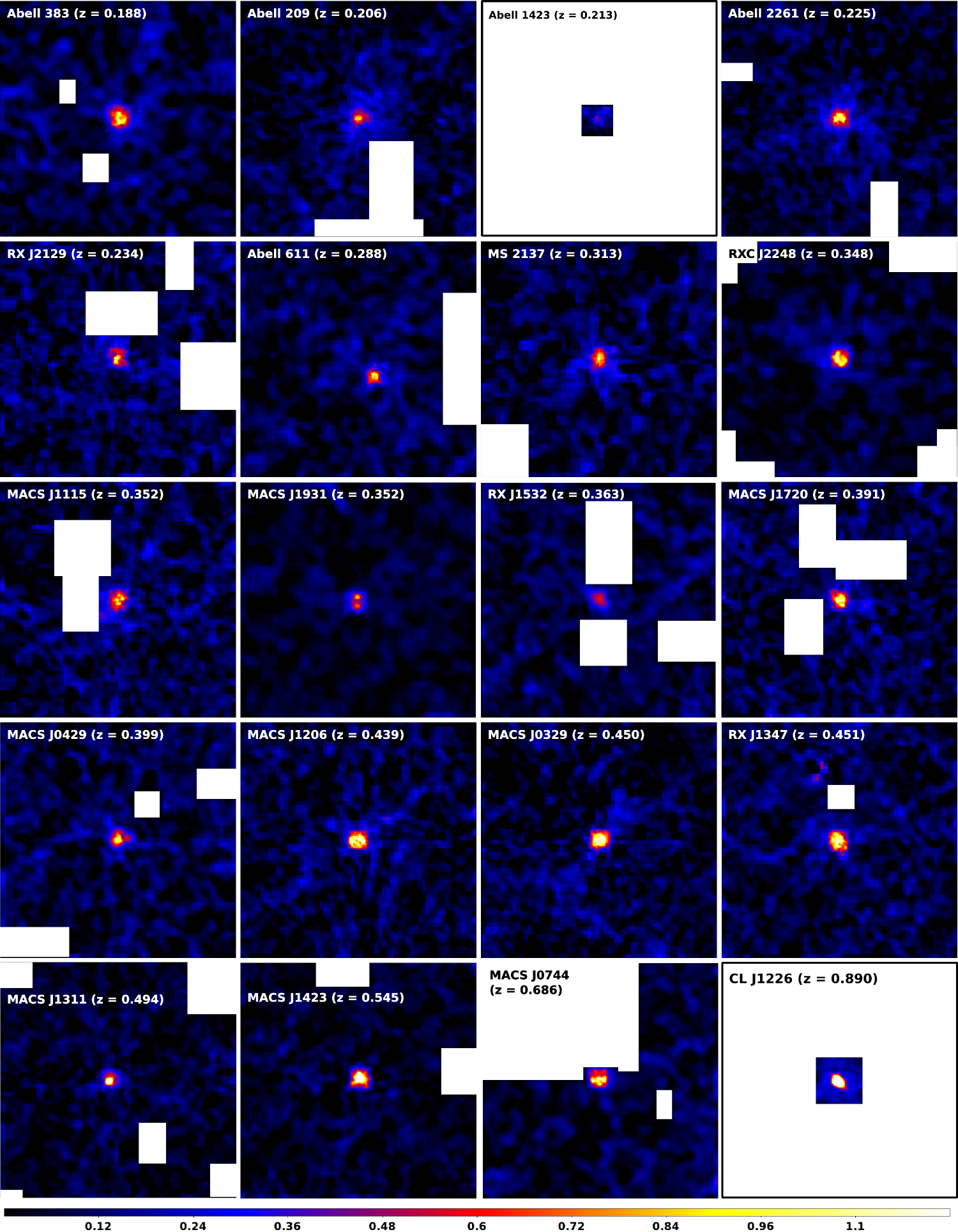}
 \end{center}
 \caption{Convergence maps for 20 X-ray selected CLASH clusters. The field size for the map of Abell~1423 is $200\arcsec$, for
 CL ~J1226 it is $300\arcsec$ and for Abell~611 it is $1400\arcsec$. For all other clusters the field size is $1500\arcsec$. The 
 color coding, together with the colorbar shows the lensing convergence, scaled to a redshift of $z=20000$. Extended white patches 
 in the convergence maps indicate field masks, usually at the position of bright foreground stars.\label{convergence_maps}}
\end{figure*}

\begin{figure*}
 \begin{center}
 \includegraphics[width=.9\textwidth]{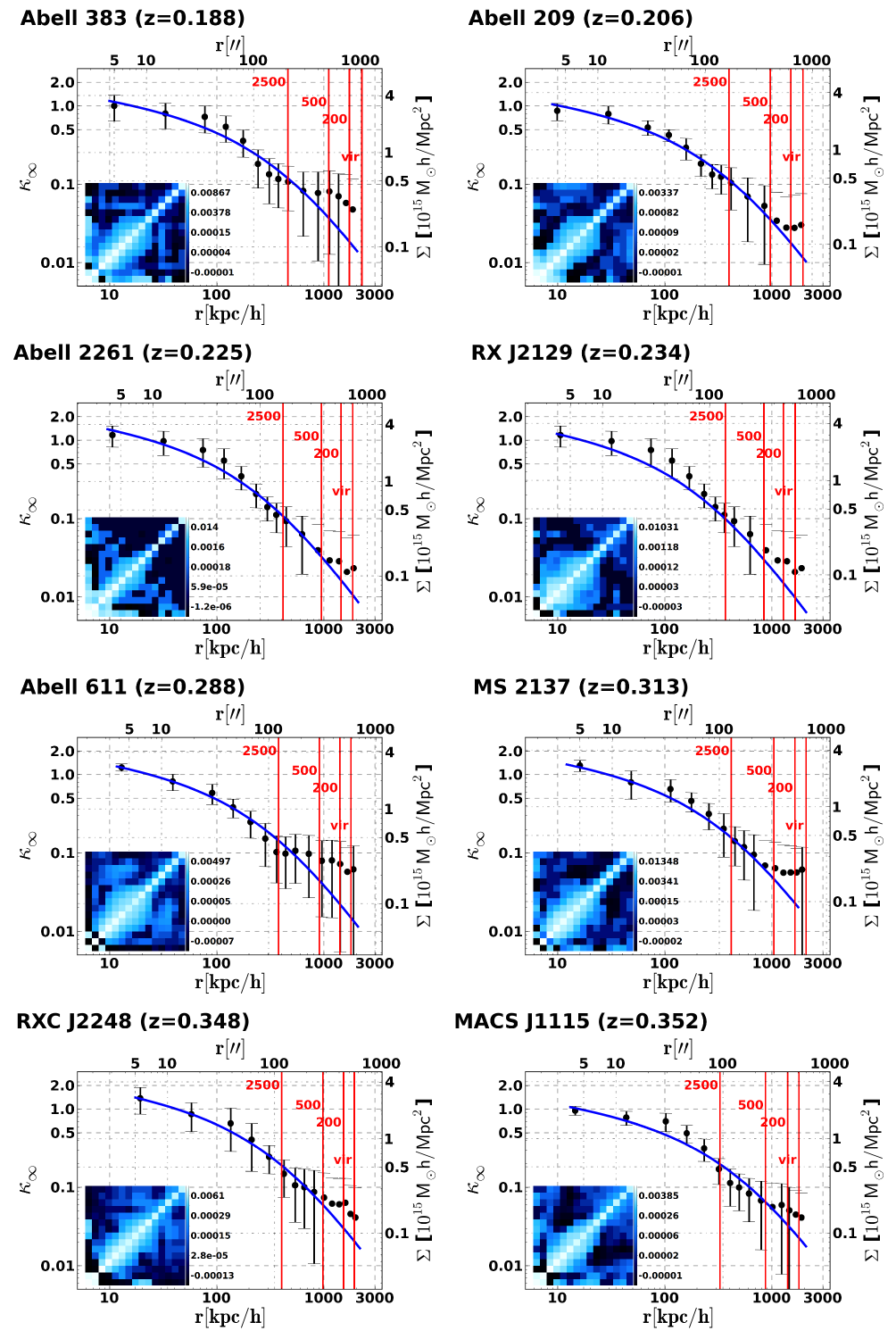}
 \end{center}
\end{figure*}

\begin{figure*}
 \begin{center}
 \includegraphics[width=.9\textwidth]{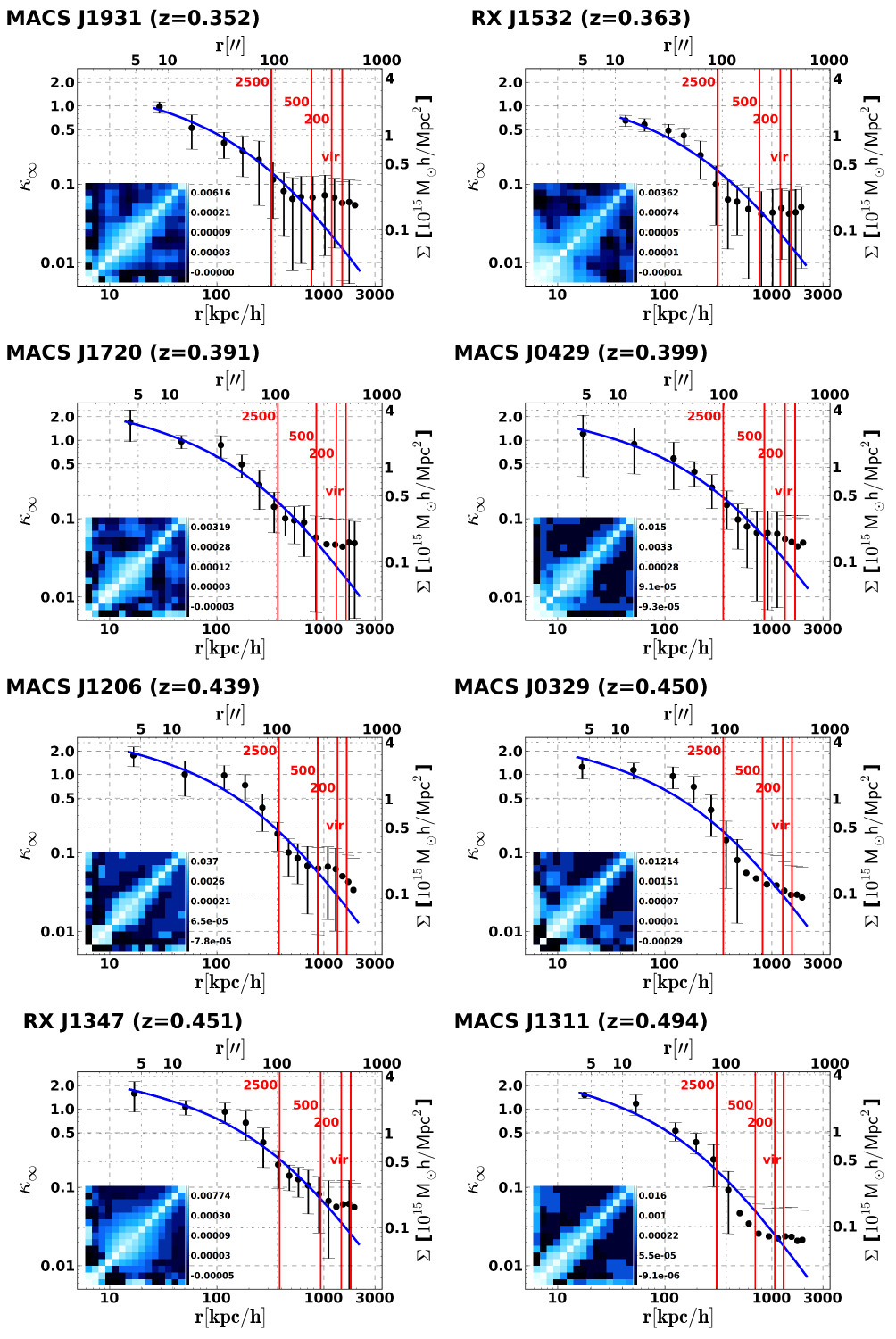}
\end{center}
\end{figure*}

\begin{figure*}
 \begin{center}
 \includegraphics[width=.9\textwidth]{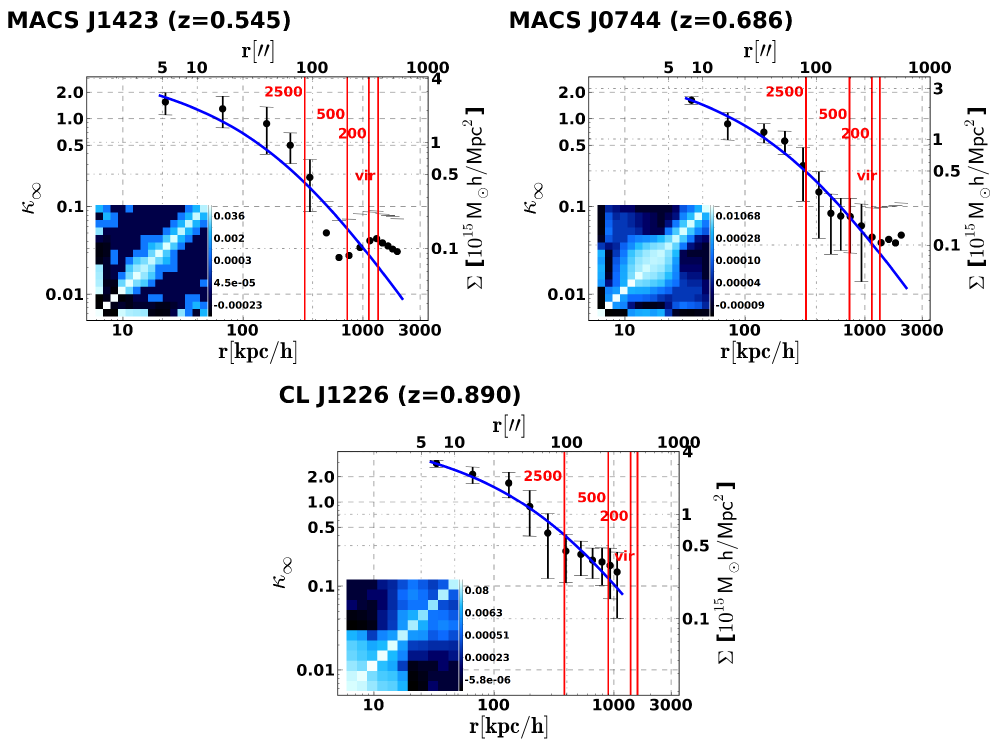}
 \end{center}
 \caption{Convergence/surface-mass density profiles for 19 X-ray selected CLASH clusters. The black data points show the mean convergence in each bin. 
  The square inset in the bottom left of each panel is the covariance matrix of the binned data and the error bars
  attached to each black data point show the square root of the diagonal elements of this matrix. Shown by the blue line
  is the best-fit NFW profile to the data. All radii refer to the peak in the dark matter density distribution
  of each halo as a center. Drawn in red are $r_{2500}$, $r_{500}$, $r_{200}$ and the virial radius of the halo.
  The convergence values are scaled to a source redshift of $z=20000$. \label{convergence_profiles}}
\end{figure*}

\end{document}